\documentstyle[prd,aps,preprint,floats,eqsecnum,tighten,epsf]{revtex} 
\begin{document}
\newcommand{\RR}{\hbox{$I$\kern-3.8pt $R$}} 
\newcommand{\rr}{\hbox{$\scriptstyle I$\kern-2.4pt $\scriptstyle R$}} 
\newcommand{\be}{\begin{equation}}
\newcommand{\ee}{\end{equation}}
\newcommand{\bea}{\begin{eqnarray}} 
\newcommand{\eea}{\end{eqnarray}} 
\newcommand{\sss}{\scriptscriptstyle}  
\newcommand{\barT}{{\bar{\cal T}}}
\newcommand{\barSigma}{{\bar\Sigma}}
\newcommand{\barn}{{\bar n}} 
\newcommand{\baru}{{\bar u}} 
\newcommand{\bara}{{\bar a}}
\newcommand{\barb}{{\bar b}}
\newcommand{\barh}{{\bar h}} 
\newcommand{\bark}{{\bar k}}
\newcommand{\bars}{{\bar s}} 
\newcommand{\bart}{{\bar t}} 
\newcommand{\barell}{{\bar\ell}}
\newcommand{\bareta}{{\bar\eta}} 
\newcommand{\barzeta}{{\bar\zeta}}
\newcommand{\barK}{{\bar K}}
\newcommand{\barP}{{\bar P}}
\newcommand{\barPi}{{\bar\Pi}}
\newcommand{\barTheta}{{\bar\Theta}}
\newcommand{\bargam}{{\bar\gamma}}
\newcommand{\barD}{{\bar{\cal D}}}
\newcommand{\epsi}{\varepsilon}
\newcommand{\jperp}{\jmath_\vdash}
\newcommand{\barj}{{\bar\jmath}}
\newcommand{\barN}{{\bar N}}
\newcommand{\barV}{{\bar V}}
\newcommand{\barM}{{\bar M}}
\newcommand{\barW}{{\bar W}}
\newcommand{\barepsi}{{\bar\varepsilon}}
\newcommand{\barjperp}{{{\bar\jmath}_\vdash}}
\newcommand{\referenceE}{{\mathsf{E}}}
\newcommand{\M}{{\cal M}} 
\newcommand{\T}{{\cal T}} 
\newcommand{\C}{{\cal C}} 
\newcommand{\U}{{\cal U}} 
\newcommand{\R}{{\Re}} 
\newcommand{\D}{{\cal D}} 
\renewcommand{\H}{{\cal H}} 
\title{Action and Energy of the Gravitational Field}  
\author{J.~D.~Brown,${}^{2}$ S.~R.~Lau,${}^{1,3}$ 
and J.~W.~York${}^{4,5}$} 
\address{\hspace{1cm}\\
${}^{1}$Raman Research Institute, Bangalore 560 080, India
\\\hspace{1cm}\\
${}^{2}$Department of Physics,\\
North Carolina State University, Raleigh, NC 27695-8202 USA   
\\\hspace{1cm}\\
${}^{3}$Applied Mathematics Group, 
Department of Mathematics,\\
University of North Carolina,  
Chapel Hill, NC 27599-3250 USA
\\\hspace{1cm}\\
${}^{4}$Theoretical Astrophysics \& Relativity Group, 
Department of Physics \&\\ Astronomy,
University of North Carolina, 
Chapel Hill, NC 27599-3255 USA
\\\hspace{1cm}\\
${}^{5}$604 Space Sciences,\\
Cornell University, Ithaca NY 14853 USA}
\maketitle 
\begin{abstract}
We present a detailed examination of the variational 
principle for metric general relativity as applied to a 
``quasilocal'' spacetime region $\M$ (that is, a region that  
is both spatially and temporally bounded). 
Our analysis relies on the Hamiltonian formulation 
of general relativity, and thereby  
assumes a foliation of $\M$ into spacelike 
hypersurfaces $\Sigma$. We allow for near complete generality in 
the choice of foliation. Using a 
field--theoretic generalization of Hamilton--Jacobi theory, 
we define the quasilocal stress-energy-momentum of the gravitational 
field by varying the action with respect to the metric on 
the boundary $\partial\M$. The gravitational stress-energy-momentum
is defined for a two--surface $B$ spanned by a spacelike hypersurface 
in spacetime. We examine the behavior of the 
gravitational stress-energy-momentum under boosts of the spanning 
hypersurface. The boost 
relations are derived from the geometrical and 
invariance properties of the gravitational action and Hamiltonian. 
Finally, we present several new examples of quasilocal 
energy--momentum, including a novel discussion of quasilocal 
energy--momentum in the large-sphere limit towards spatial infinity.
\vfill
\noindent
Chapel Hill and Raleigh, April 2002
\end{abstract}
\newpage
\section{Introduction}

Beginning with the earliest days of general relativity 
and continuing to the present, relativists have actively
sought to define gravitational stress-energy-momentum 
({\sc sem}) from a variational principle. The motivation
to do so is readily apparent. {\sc sem}, and energy in 
particular, plays a central role in most branches of 
physics. In this paper we discuss a relatively new approach (see 
for example${}^1$
Refs.~\cite{BY,HW1,Hayward1,Lau0,LauDiss,BrHy,Martinez,BCM,Lau1,Hayward2,FM,Lau2,Lau3,HH1,HH2,BLY1,CECpaper,BM1,BLY2,Dadhich,Lau4,BM2,BC,Epp,AncoTung})
to the problem, which we refer to here 
as the canonical quasilocal formalism ({\sc cqf}). The
{\sc cqf} is based upon a field--theoretic generalization 
of Hamilton--Jacobi theory. 
We present many results new to the {\sc cqf} and, in the process, 
recover the recent results from Refs.~\cite{Lau2,BM1}. 

Over the last thirty years, research has yielded a 
more--or--less satisfactory understanding of total 
energy--momentum for asymptotically flat spacetimes and 
asymptotically anti-de Sitter spacetimes. However, as 
no physical system is ever truly isolated, these asymptotic 
conditions---however useful---are ultimately unphysical, 
theoretical idealizations. In any case, practical numerical 
calculations are always restricted to a spatially finite 
region. For this reason and others (see the next paragraph), 
recent efforts have turned to the issue of defining {\sc sem} 
quasilocally, that is to say, associating gravitational
{\sc sem} with spatially bounded regions. (Indeed, one direct
application of the formalism we present here is an approach to 
numerical outer boundary conditions for the gravitational field
described in a forthcoming paper.\cite{ArrestedHistory})

As we will see, the {\sc cqf} naturally leads to a 
definition of gravitational {\sc sem} that is quasilocal. 
We are motivated primarily by the desire to obtain physically
meaningful and useful energy--like quantities that characterize 
the {\em classical} gravitational field in a bounded region. 
However, our original motivation for developing the {\sc cqf} 
stemmed from a problem in {\em semiclassical} gravity, namely, 
understanding thermodynamical internal energy for black holes. 
The asymptotically--defined Arnowitt-Deser-Misner ({\sc adm})
energy\cite{ADM}, for example, cannot serve as a useful
internal energy because an infinite, self-gravitating system 
at finite temperature is thermodynamically unstable. Thus, 
the partition function can be defined only for systems with 
finite spatial extent, and this necessitates a quasilocal 
definition of energy (see
Refs~\cite{BBWY,BMY,BY2,BCM,Brown1,Brown2,Brown3,Brown4} 
and references therein).

Before turning to the {\sc cqf}, let us mention several 
approaches toward defining gravitational energy from a variational 
principle. The history of this problem is long, so an encompassing 
study would require a separate, extensive review. Here, we give 
only a brief summary of several of the historically important works. 
These works are based on a field--theoretic generalization of 
Noether's theorem \cite{Noether}. 

Einstein was the first to derive gravitational {\sc sem} 
from an action principle.\cite{Einstein} By discarding a 
metric--dependent divergence term in the second--order 
covariant Hilbert action, he obtained a first--order action, 
the so--called $\Gamma\Gamma$ action, that is the 
four--integral of a bulk Lagrangian quadratic in the 
Christoffel symbols. He then carried out a Noether--type 
analysis, and derived a canonical gravitational 
{\sc sem} {\em pseudo}tensor and its corresponding 
super--potential.${}^{2}$ Given what we've learned about 
the asymptotic structure of spacetime in the decades 
since this early work, it is remarkable how successful 
the Einstein definitions were.\cite{GoldberginHeld} Most 
of the key properties of spatial infinity (including 
decay of the metric and derivatives of the metric) 
are found in Einstein's original paper. The drawback of 
Einstein's approach is that the $\Gamma\Gamma$ action is 
not fully diffeomorphism invariant (it is invariant 
modulo boundary terms), and his gravitational 
{\sc sem} is coordinate dependent. At the quasilocal level 
there is no obvious general prescription for how one 
should choose coordinates.

In the early 1960's M\o ller discovered a new bulk action
for general relativity that is similar to Einstein's $\Gamma\Gamma$ 
action, but is quadratic in the tetrad connection (Ricci 
rotation) coefficients.\cite{Moller} We might refer to it 
as the $\omega\omega$ action. Like the Einstein Lagrangian, 
the M\o ller bulk Lagrangian differs from the Hilbert 
Lagrangian by a pure divergence. (See also related work
in Refs.~\cite{Nelson,Goldberg,Sparling}.) Although the 
M\o ller action is 
not fully invariant under ``internal'' transformations of 
the tetrad, it is diffeomorphism invariant and is therefore 
arguably preferable to the Einstein action as the starting 
point for a Noether--type analysis. Moreover, the resulting 
theory of {\sc sem} can be translated readily into the 
language of two--spinors, a powerful formalism which has 
led to numerous results in the quasilocal setting (see, 
for instance, the works of Szabados,
Refs.~\cite{Szabados0,Szabados1,Szabados2,Szabados3} and 
references therein). We point out, however, that in adopting 
the M\o ller action, one is departing from pure 
metric relativity, Einstein's original theory. It 
is not clear at all that the {\sc sem} concepts derived 
in any such framework ``pull back'' to the metric phase 
space.

By introducing a background metrical structure, one may 
isolate---in a coordinate independent fashion---a purely 
metric divergence term in the Hilbert action. In 
Ref.~\cite{Rosen} Rosen discarded such a term, thereby
obtaining another bulk action amenable to Noether techniques. 
An invocation of the Noether theorem in purely metric gravity, 
this approach towards defining gravitational {\sc sem} is 
close in spirit to Einstein's, and may be considered as a 
a refined version of his original analysis. However, 
the approach would seem limited in that there is not always 
a natural choice of background spacetime. Bi\v c\'ak, 
Lynden-Bell, Katz, and Petrov have developed and used an improved 
version of the approach in several recent papers (see 
Refs.~\cite{KBL,PK,Petrov} and references therein) addressing,
among other things, gravitational perturbations of 
cosmological solutions to 
the Einstein equations.

We now turn to the canonical quasilocal formalism. Our 
analysis is based on the so-called ``Trace-K'' 
action\cite{York,BY,HW1,Hayward1}, 
which differs from the standard 
Hilbert action by metric-dependent boundary terms. Its use 
leads to a purely metric formalism, as does the Einstein 
$\Gamma\Gamma$ action. However, unlike the $\Gamma\Gamma$ 
action, the Trace-K action${}^{3}$ 
is manifestly invariant under coordinate 
transformations. Moreover, the Trace-K action does not
depend on any background structures.${}^{4}$
Since the Trace-K action does not stem from 
a bulk Lagrangian, it is not immediately clear how to 
apply the Noether theorem. But we can bypass a Noether 
analysis altogether, using instead the {\sc cqf} which 
is based on Hamilton-Jacobi theory. We point out that our 
approach is intimately related to a body of work done by 
Kijowski and co-workers (see Ref.~\cite{Kijowski} and references 
therein). Kijowski's approach starts with 
novel and important ideas from symplectic theory \cite{KT}, and examines 
the relevant symplectic geometry in great detail. Our approach, 
on the other hand, 
starts with standard Hamilton-Jacobi theory and focuses primarily on 
the physical spacetime geometry. We stress 
that both approaches are merely different faces of a Hamiltonian 
analysis, and thus somewhat different from more 
traditional approaches based on Noether techniques.

Consider a spatially and temporally bounded spacetime region $\M$ 
with metric $g_{\mu\nu}$ and
boundary $\partial\M$. The boundary $\partial\M$ of the region consists of
a timelike element $\barT$ (the meaning of the ``bar'' is explained below)
and spacelike elements $\Sigma'$ and $\Sigma''$. Such a spacetime region
is depicted in Fig.~1, but note that $\barT$ need not be
connected.
\begin{figure}[!htb]      
\epsfbox{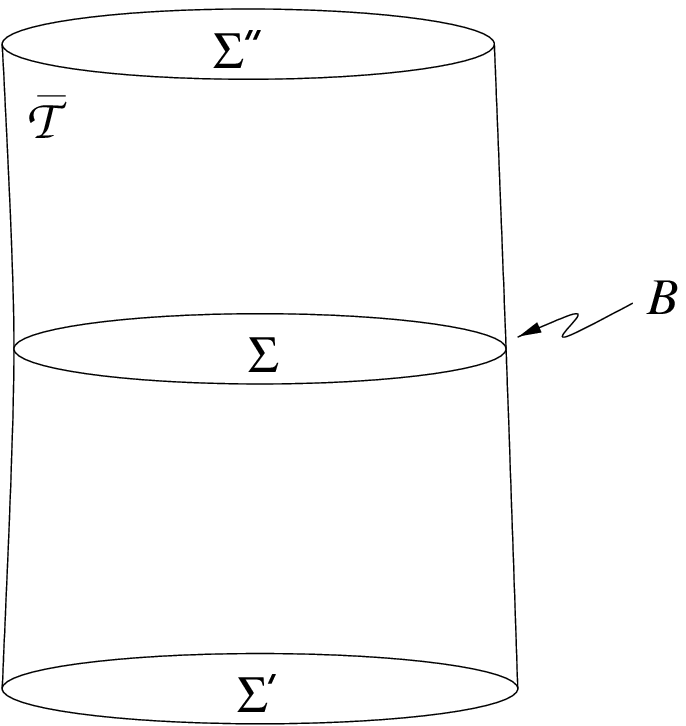}     
\caption{}
\end{figure} 
We assume that the spacetime $\M$ is
foliated into spacelike hypersurfaces $\Sigma$, defined by 
$t = {\rm const}$. Further, we require
the boundary of each $\Sigma $ leaf to lie in $\barT$.${}^{5}$
The intersections of the leaves of the spacetime foliation $\Sigma$
with the timelike boundary element $\barT$ define a foliation of
$\barT$ into two--dimensional spacelike hypersurfaces $B$ (which need not
be connected). The region ${\cal M}$ may itself be contained
in some ambient spacetime. We note that the boundary $B$ and its history $\barT$ 
are simply submanifolds of spacetime and need not be
physical barriers. 

The future--pointing unit normal to the $t = {\rm const}$ 
hypersurfaces is denoted $u^\mu$, and the outward--pointing 
unit normal of $\barT$ is denoted $\barn^\mu$. 
The induced metric on $\Sigma$ is $h_{ij}$ and 
the induced metric on $\barT$ is $\bargam_{ij}$. Because 
in general $\barn^{\mu} u_\mu \neq 0$ on $\barT$, the Eulerian 
observers of the $B$ foliation of $\barT$, comoving with 
$\barT$, need not be at rest with respect to the $\Sigma$ slices. 
This necessitates the ``barred'' and ``unbarred'' notation which
keeps track of the two sets of Eulerian observers at the boundary:
those comoving with $\barT$ and at rest with respect to the $B$
foliation of $\barT$ (the ``barred observers") and those at rest 
with respect to the $\Sigma$ foliation of $\M$ (the ``unbarred 
observers"). 

The four--velocities of the barred observers will 
be denoted ${\bar u}^\mu$. Similarly, at each point of $\barT$, we define 
$n^\mu$ as the unit outward pointing vector for the unbarred observers. 
That is, $n^\mu$ lies in $\Sigma$ and 
is orthogonal to $B$.${}^{6}$
Note that, by construction, ${\bar u}^\mu {\bar n}_\mu = 0$ and 
$u^\mu n_\mu = 0$. These vectors are related by the boost relations 
\bea 
\barn_\mu &=& \gamma n_\mu + \gamma v u_\mu \ ,
\label{barrelations}\eqnum{\ref{barrelations}a}\\
\baru_\mu &=& \gamma u_\mu + \gamma v n_\mu \ .
\eqnum{\ref{barrelations}b} 
\addtocounter{equation}{1}
\eea
where $v$ is the boost velocity between the two sets of observers 
and $\gamma = (1-v^2)^{-1/2}$. In Appendix A we present the details 
of the kinematical relationships needed for this paper. 

As mentioned, our analysis is carried out in the purely 
metric formulation of gravity and is based on the 
Trace K action,\cite{York,HW1,Hayward1,CECpaper}
\be 
S[g] = \frac{1}{2\kappa} \int_{\M}d^4x\sqrt{-g}\,{\R} 
+ \frac{1}{\kappa} \int_{\Sigma'}^{\Sigma''}d^3x \sqrt{h}\,K 
- \frac{1}{\kappa} \int_{\barT} d^3x \sqrt{-\bargam}\,
  \barTheta 
- \frac{1}{\kappa} \int^{B''}_{B'} d^{2}x\sqrt{\sigma}
  \theta
\ . \label{TraceKaction} 
\ee
Here, $\kappa$ is $8\pi$ times Newton's constant. For simplicity we have 
omitted matter and cosmological constant contributions to the action. The 
symbol $\int_{\Sigma'}^{\Sigma''}$ is shorthand for $\int_{\Sigma''} 
- \int_{\Sigma'}$, and $K$ is the trace of the extrinsic 
curvature $K_{\mu\nu} = - h_\mu^\alpha \nabla_\alpha u_\nu$
of the boundary elements $\Sigma'$ and $\Sigma''$.
Similarly, the function $\barTheta$ is the 
trace of the extrinsic curvature $\barTheta_{\mu\nu} = 
-\bargam_\mu^\alpha \nabla_\alpha \barn_\nu$
of the boundary element $\barT$.  
The action (\ref{TraceKaction}) includes contributions (first considered 
in Refs.~\cite{HW1,Hayward1}) from the ``corners" 
$B' = \Sigma'\cap \barT$ and $B'' = \Sigma'\cap \barT$, where 
$\sigma$ is the determinant of the metric $\sigma_{ab}$ on the 
corners. The velocity parameter $\theta$ is defined by 
$\sinh\theta = -u^\mu \barn_\mu = \gamma v$. 

The Trace-K action
has the crucial property that its associated
variational principle features fixation of the induced three
metric ${}^{3}g_{ij}$ on $\partial\M$.${}^{7}$
In particular, the lapse of proper time for an observer, comoving with
$\barT$ and at rest with the $B$ foliation, is fixed as boundary data since this
information is encoded in the fixed $\bar{\cal T}$ three--metric.
The value of the {\em quasilocal energy surface density} (at  
a given point on the observer's worldline) is defined 
through a {\sc hj}  variation as minus the rate of change of the 
classical action with respect to an
infinitesimal stretch (enacted at the given point) in the proper
time separation between $\Sigma'$ and $\Sigma''$.
Of course, the $\bar{\cal T}$
three--metric specifies more than just the lapse of proper time
between the initial and final slices---it contains information
about all possible spacetime intervals on $\bar{\cal T}$.
One is free to consider the changes in the classical action
corresponding to arbitrary {\sc hj} variations in the
$\bar{\cal T}$ metric. The original Ref. \cite{BY} has
demonstrated how this freedom leads not only to the 
energy surface density but also to surface densities
for {\em tangential momentum} and {\em spatial stress}
(both are pointwise--defined $B$ tensors). 

In this paper, we 
extend the {\sc cqf} analysis of Ref.\cite{BY} by considering changes 
in the classical action corresponding to {\sc hj} variations in
$\Sigma''$ 
(or $\Sigma'$) boundary data. This leads to quasilocal surface
densities for {\em normal momentum}, {\em tangential
momentum} (which is equivalent to the previous definition), and
{\em temporal stress} (this was also shown in Ref.~\cite{Lau2}). 
Therefore, the quasilocal stress-energy-momentum consists of energy,
normal momentum, and tangential momentum surface densities and spatial 
and temporal stress tensors.

We also extend the {\sc cqf} by considering ``boost relations" 
between the quasilocal surface densities as defined by barred and 
unbarred observers. These {\sc sem} boost relations can be viewed 
as canonical realizations of the relations (\ref{barrelations}) 
satisfied by the barred and unbarred observers' unit vectors. 
The observer dependence of the quasilocal 
{\sc sem} is best described from the following perspective. The 
various {\sc sem} quantities are defined as tensors on the spatial 
boundary $B$ spanned by a spacelike hypersurface $\Sigma$. The 
boost relations characterize the behavior of these tensors under 
a boost of the spanning slice $\Sigma$; that is, they characterize 
the dependence of the quasilocal {\sc sem} on the choice of observers 
passing through $B$. As purely geometrical relations, the boost
relations among energy, tangential momentum, and normal momentum 
surface densities have been noted elsewhere in the literature 
(see, for instance, Refs.~\cite{Szabados0,Kijowski}). Moreover, 
their particular role in the {\sc cqf} has been pointed out in 
Refs.~\cite{LauDiss,Lau2}. Here we present a unique 
derivation of these relations, demonstrating that their 
geometrical content is already encoded in the 
gravitational Hamiltonian.

Finally, we present several new examples of quasilocal energy--momentum, 
including an analysis of cylindrical gravitational waves. We also include 
a novel discussion of quasilocal energy--momentum in the large-sphere limit 
towards spatial infinity. Agreement between the Trautman--Bondi--Sachs 
total graviational energy--momentum and the notion of quasilocal 
energy--momentum arising in the {\sc cqf} has been considered 
elsewhere.\cite{BLY1}

In Sec.~II we prepare for the Hamilton--Jacobi variation of the 
Trace-K action by considering the general variation of the 
Hilbert action. Of particular importance are the corner terms 
that arise at the intersections of $\bar{\cal T}$ with $\Sigma'$ 
and $\Sigma''$. These terms have appeared previously in the 
literature.\cite{HW1,Hayward1,Lau2,Kijowski} However, to our
knowledge, Sec.~II contains the first completely geometrical 
derivation of the result (although the same result was 
obtained explicitly via another method in Ref.~\cite{Kijowski}). 
In Sec.~III we apply the {\sc cqf} to the Trace-K action 
and derive the quasilocal {\sc sem}. In the process, 
we obtain the boost relations among the energy and momentum 
surface densities and spatial and temporal stress tensors 
as defined by the boosted and unboosted observers. We also
discuss the notion of boost invariants which
allow for the construction of several mass definitions which
have appeared in the relativity literature.
Section IV contains a derivation 
of the Hamiltonian form of the action and its variation. 
We then derive the boost relations for the energy and 
momentum surface densities by boosting the gravitational 
Hamiltonian. Finally, Sec.~V and Sec.~VI deal with concrete
examples and applications, and in these sections we chose 
units such that the constant $\kappa$ appearing in 
Eq.~(\ref{TraceKaction}) is simply $8\pi$. In Sec.~V we first
deal with the issue of zero--points for the quasilocal densities,
and the relationship between these zero-points and a freedom
always present in {\em any} variational principle. The freedom
concerns appending to the action, here the gravitational Trace-K action, 
an arbitrary functional of the fixed boundary data.
Having (at least partially) dealt with this issue, we then
write down quasilocal energy and momentum expressions for a variety 
of exact solutions to the Einstein equations; and in Sec.~VI we apply 
our formalism to spacetimes with are asymptotically flat in
spacelike directions, in the process making some novel
observations about total gravitational energy at spatial
infinity. 

Appendix A contains several key 
kinematical results that are used throughout the paper. 
Appendix B is devoted to the derivation 
of certain curvature splittings needed for the analysis 
in Sec.~III. Finally, in Appendix C, we show that the 
rate of change of the boost parameter equals the 
normal gradient of the lapse function defining the boost. This 
is needed for the analysis in Sec.~IV. 

\section{Variation of the Hilbert action}
In this section we consider the standard Hilbert 
action,\cite{MTW}
\begin{equation}
       S_{\scriptscriptstyle H}[g] 
     = \frac{1}{2\kappa}\int_{\cal M}
       d^{4}x \sqrt{-g}{\,} \Re {\,},
\label{Einstein_Hilbert_action}
\end{equation}
and its associated variational principle as applied to
a bounded spacetime region ${\cal M}$, a careful analysis
of which is crucial for the entire discussion. Such
an analysis is, of course, not 
new\cite{York,HW1,Hayward1,CECpaper,BM1,Kijowski}; 
however, as we do give a new version of a nontrivial
calculation of fundamental importance, we believe the details 
belong up front and not relegated to an appendix.

The relevant geometry of the various foliations of 
${\cal M}$ is described in the Introduction and Appendix A. 
We examine the variation 
$\delta S_{\scriptscriptstyle H}$ 
of the action induced by an infinitesimal variation 
$\delta g_{\mu\nu}$ in the metric tensor and derive
the following result:\cite{York,HW1,Hayward1,CECpaper,BM1,Kijowski}
\begin{equation}
\delta S_{\scriptscriptstyle H} =  
     - \frac{1}{2\kappa} \int_{\cal M}d^{4} x
       \sqrt{-g} G^{\mu\nu} \delta g_{\mu\nu}
     - \int^{\Sigma''}_{\Sigma'} 
       d^{3} x h_{ij} \delta P^{ij}
     - \int_{\bar{\cal T}}
       d^{3} x \bar{\gamma}_{ij} 
       \delta \bar{\Pi}{}^{ij} 
     + \frac{1}{\kappa} \int^{B''}_{B'} d^{2}x
       \sqrt{\sigma} \delta \theta {\,} .
\label{mainresult}
\end{equation}
In this expression
$G^{\mu\nu}$ is the Einstein tensor, $\theta$ is the 
velocity parameter described earlier, and
\begin{eqnarray} 
P^{ij} & = & \frac{\sqrt{h}}{2\kappa} 
\left(K h^{ij} - K^{ij}\right)
\label{momenta} \eqnum{\ref{momenta}a} \\
\bar{\Pi}^{ij} & = & - \frac{\sqrt{-\bar{\gamma}}}{2\kappa}
\left(\bar{\Theta} \bar{\gamma}{}^{ij} - \bar{\Theta}{}^{ij}
\right) 
\eqnum{\ref{momenta}b}
\addtocounter{equation}{1}
\end{eqnarray}
are respectively the $\Sigma$ and $\barT$ gravitational 
momenta. 

In writing the Hilbert action (\ref{Einstein_Hilbert_action}) 
and its variation (\ref{mainresult}), we do not necessarily assume that 
the spacetime boundary elements $\Sigma'$, $\Sigma''$, and $\barT$ have 
smooth embeddings in $\M$. That is, the unit normal ${\bar n}^\mu$ of 
$\barT$ and the unit normals $u^\mu$ of $\Sigma'$ and  $\Sigma''$ need 
not be continuous vector fields. For example, the timelike boundary 
element $\barT$ can contain a ``kink", a spacelike two-surface at which 
${\bar n}^\mu$ changes discontinuously. In that case the $\barT$ boundary 
term in $\delta S_{\scriptscriptstyle H}$ contains a contribution from 
the kink. The form of the contribution is discussed in
Ref.~\cite{BrHy}.

\subsection{Preliminary results and a lemma}

To begin, let us collect a few results concerning 
the variation $\delta \Gamma^{\lambda}{}_{\mu\nu}$ of the 
affine connection induced by an infinitesimal 
$\delta g_{\mu\nu}$. With these
we prove a lemma of particular use when examining the 
variation $\delta S_{\scriptscriptstyle H}$ of the 
action.${}^{8}$
First, expansion of the identity 
$\delta (\nabla_{\lambda} g_{\mu\nu}) = 0$ leads directly 
to the result
\begin{equation}
       2\delta \Gamma_{(\mu\nu)\lambda} 
     = \nabla_{\lambda} \delta g_{\mu\nu} {\,}.
\label{deltaGamma1}
\end{equation}
Second, the well--known formula\cite{MTW} 
$\delta \Re_{\mu\nu} 
= \nabla_{\lambda} \delta \Gamma^{\lambda}{}_{\mu\nu}
- \nabla_{\nu} \delta \Gamma^{\lambda}{}_{\mu\lambda}$ for 
the induced variation of the Ricci tensor implies that the 
contracted variation $g^{\mu\nu} \delta \Re_{\mu\nu}$ is a
pure spacetime divergence. Indeed, writing 
$g^{\mu\nu} \delta \Re_{\mu\nu} = \nabla_{\mu} V^{\mu}$,
we find that
\begin{equation}
       V^{\mu} = 2\delta \Gamma^{[\mu\nu]}{}_{\nu} {\,}.
\label{deltaGamma2}
\end{equation}
Finally, consider a metric--dependent covector 
$\omega_{\mu}$ (i.~e.~$\delta \omega_{\mu}$ need
not vanish) and the spacetime divergence 
$\nabla_{\mu} \omega^{\mu}$ constructed from it. The
variation of this divergence is
\begin{equation}
\delta \nabla_{\mu} \omega^{\mu} = 
     - {\textstyle \frac{1}{2}}\omega_{\mu} V^{\mu}
     + \nabla^{\mu} \delta \omega_{\mu} 
     - {\textstyle \frac{1}{2}}
       (\nabla^{\mu} \omega^{\nu})\delta g_{\mu\nu} 
     - {\textstyle \frac{1}{2}} \nabla^{\mu}(\omega^{\nu}
       \delta g_{\mu\nu}) {\,} .
\label{deltaGamma3}
\end{equation}
To obtain this result, expand the variation 
$\delta (g^{\mu\nu}\nabla_{\mu} \omega_{\nu})$, insert
the identity 
$\delta \nabla_{\mu}\omega_{\nu} = \nabla_{\mu} 
\delta \omega_{\nu} - \omega_{\lambda} 
\delta\Gamma^{\lambda}{}_{\mu\nu}$, and then use 
Eqs.~(\ref{deltaGamma1}) and (\ref{deltaGamma2}). 

We now prove the following.\\
\underline{{\sc lemma}}: Consider a unit 
{\em hypersurface--orthogonal} vector field, say $u^{\mu}$
with normalization $u^{\mu} u_{\mu} = \epsilon$ (with 
$\epsilon = \pm 1$).${}^{9}$
Also consider the induced
metric $h_{\mu\nu} = g_{\mu\nu} - \epsilon u_{\mu} u_{\nu}$ 
on the hypersurfaces to which $u^{\mu}$ is orthogonal, as
well as the extrinsic curvature tensor 
$K_{\mu\nu} = - h^{\alpha}_{\mu} \nabla_{\alpha} u_{\nu}$.
The variation $\delta K$ of the mean 
curvature $K = - \nabla_{\mu} u^{\mu}$ satisfies the equation
\begin{equation}
2\delta K = 
       u_{\mu} V^{\mu}
     - K^{\mu\nu} \delta h_{\mu\nu}
     - D_{\alpha} (h^{\alpha}_{\mu} \delta u^{\mu}){\,} ,
\label{lemma}
\end{equation}
where $D_{\alpha}$ is the covariant
derivative operator compatible with $h_{\mu\nu}$.

Our proof of the lemma makes use of the following two
identities: (i)
$\nabla^{\mu} u^{\nu} = - K^{\mu\nu} + \epsilon u^{\mu} 
u^{\lambda}\nabla_{\lambda} u^{\nu}$ and (ii) 
$\delta u_{\mu} =  {\textstyle \frac{1}{2}} \epsilon u_{\mu}
u^{\alpha} u^{\beta} \delta g_{\alpha\beta}$. Identity
(i) follows directly from the definition of $K_{\mu\nu}$ 
and the spacetime expression for the induced metric
$h_{\mu\nu}$. We verify (ii) by writing $u_{\mu} = \epsilon
N \nabla_{\mu} t$, where the coordinate $t$ labels the 
hypersurfaces to which $u_{\mu}$ is orthogonal and 
$N = (\epsilon \nabla_{\mu}t{\,} g^{\mu\nu}{\,} 
\nabla_{\nu}t)^{-1/2}$ is the lapse function. 
As the first step towards proving the lemma, we rewrite
Eq.~(\ref{deltaGamma3}) with $u_{\mu}$ 
in place of $\omega_{\mu}$,
make substitutions with the identities (i) and (ii), and do
a bit of algebra in order to obtain
\begin{equation}
2\delta K = 
       u_{\mu} V^{\mu}
     - K^{\mu\nu} \delta g_{\mu\nu}
     - \epsilon (\nabla_{\lambda} u^{\lambda})
        u^{\mu} u^{\nu} \delta g_{\mu\nu}
     + h^{\mu\lambda}\nabla_{\lambda}( 
       u^{\nu}\delta g_{\mu\nu}) {\,}.
\label{lemmastep1}
\end{equation}
Now, by (i)  
$\nabla_{\mu} u^{\mu} = h^{\mu\nu} \nabla_{\mu} u_{\nu}$; and,
therefore, we may collect the last two terms on the right--hand
side in Eq.~(\ref{lemmastep1}), thereby arriving at
\begin{equation}
2\delta K = 
       u_{\mu} V^{\mu}
     - K^{\mu\nu} \delta g_{\mu\nu}
     + h^{\lambda\kappa} \nabla_{\lambda}(h^{\mu}_{\kappa} 
       u^{\nu}\delta g_{\mu\nu}) {\,}.
\label{lemmastep2}
\end{equation}
Finally, since $K^{\mu\nu}$ is purely spatial, 
         $K^{\mu\nu} \delta g_{\mu\nu} 
        = K^{\mu\nu} \delta h_{\mu\nu}$.
Moreover, with identity (ii) we can show that 
         $h^{\mu}_{\kappa} u^{\nu} \delta g_{\mu\nu} = 
        - h_{\mu\kappa} \delta u^{\mu}$. 
Substitution of these results along with the definition of 
$D_{\alpha}$ into Eq.~(\ref{lemmastep2}) completes the proof.

As we have been careful to allow for the case $\epsilon = 1$,
our proof of the lemma establishes
\begin{equation}
2\delta \bar{\Theta} = 
       \bar{n}_{\mu} V^{\mu}
     - \bar{\Theta}^{\mu\nu} \delta \bar{\gamma}_{\mu\nu}
     - \bar{\cal D}_{\alpha} 
       (\bar{\gamma}{}^{\alpha}_{\mu} \delta 
       \bar{n}{}^{\mu})
\label{corollary}
\end{equation}
as a corollary. Here, $\bar{\cal D}_{\alpha}$ is the 
covariant derivative compatible with the $\bar{\cal T}$
metric $\bar{\gamma}_{ij}$.

\subsection{Variation of the action}

Our goal now is to obtain the expression 
(\ref{mainresult}) for the variation 
$\delta S_{\scriptscriptstyle H}$ of the 
Hilbert action. Simple manipulations 
show that the variation of the action is
\begin{equation}
\delta S_{\scriptscriptstyle H} = 
       \frac{1}{2\kappa} \int_{\cal M} d^{4} x
       \sqrt{-g} \left( - G^{\mu\nu} \delta g_{\mu\nu} 
     + \nabla_{\mu} V^{\mu}\right) {\,} ,
\label{deltaS_EH}
\end{equation}
where $V^{\mu}$ has been defined in 
Eq.~(\ref{deltaGamma2}). Focus attention on the divergence
term in Eq.~(\ref{deltaS_EH}), namely,
\begin{equation}
(\delta S_{\scriptscriptstyle H})_{\partial {\cal M}}
= \frac{1}{2\kappa} \int_{\cal M} d^{4} x
       \sqrt{-g} \nabla_{\mu} V^{\mu} {\,} .
\label{divergenceterm}
\end{equation}
Via Stokes' theorem,\cite{Wald}
\begin{equation}
       \int_{\cal M}( \nabla_{\mu}
       V^{\mu}) {\boldmath \mbox{$\epsilon$}}  
     = \int_{\partial {\cal M}}
       ( V \cdot {\boldmath \mbox{$\epsilon$}})
{\,} ,
\label{StokesTHM}
\end{equation}
we can express 
$(\delta S_{\scriptscriptstyle H})_{\partial {\cal M}}$
as a pure boundary term. In Eq.~(\ref{StokesTHM}) we have 
used abstract notation for both the spacetime volume form 
$({\boldmath \mbox{$\epsilon$}})_{\mu\nu\lambda\kappa} =
\epsilon_{\mu\nu\lambda\kappa}$ 
and the three--form 
$(V \cdot {\boldmath \mbox{$\epsilon$}})_{\nu\lambda\kappa} 
= V^{\mu} \epsilon_{\mu\nu\lambda\kappa}$. The integral on 
the right--hand side of Eq.~(\ref{StokesTHM}) can 
be written as a sum of separate integrals over each 
element of the boundary $\partial {\cal M} = 
\Sigma' \bigcup \Sigma''
\bigcup \bar{\cal T}$. Indeed, no corner-term 
contributions (i.~e.~two-surface integrals 
over $B'$ and $B''$) can arise at this stage, because
both $B'$ and $B''$ are sets of measure zero with respect to 
$\partial {\cal M}$ and the integrand 
$(V \cdot {\boldmath \mbox{$\epsilon$}})$ is continuous
(as both $V^{\mu}$ and $\epsilon_{\mu\nu\lambda\kappa}$ are 
continuous). Now, standard convention fixes the orientation of the 
boundary $\partial {\cal M}$ by choosing the outward--pointing 
normal ${\sf n}{}^{\mu}$ to $\partial {\cal M}$ as embedded in 
${\cal M}$; that is to say, the alternating tensor on 
$\partial {\cal M}$ is taken to be 
$\epsilon_{\nu\lambda\kappa} 
= {\sf n}{}^{\mu} \epsilon_{\mu\nu\lambda\kappa}$. Subject to
this convention, one finds in general that
\begin{equation}
       \int_{\partial {\cal M}}
       ( V \cdot {\boldmath \mbox{$\epsilon$}}) 
     = \int_{\partial {\cal M}}d^{3}x 
       \sqrt{|{}^{3}g|} \epsilon {\sf n}_{\mu} V^{\mu}{\,} .
\label{boundaryintegral}
\end{equation}
In this expression $|{}^3 g|$ is the (absolute value of) the 
determinant of the induced metric on $\partial {\cal M}$ and 
$\epsilon = {\sf n}_{\mu} {\sf n}^{\mu}$ is a sign factor
which is either $1$ or $-1$ depending on the boundary element. 
Notice that $\epsilon {\sf n}_{\mu}$
is the covector dual to the outward--pointing normal 
${\sf n}^{\mu}$. Therefore, for the case at hand
with 
$\partial {\cal M} = \Sigma' \bigcup \Sigma'' \bigcup \bar{\cal T}$,
we expand the right--hand side of Eq.~(\ref{boundaryintegral}), 
and obtain
\begin{equation}
       (\delta S_{\scriptscriptstyle H})_{\partial {\cal M}}  
     = - \frac{1}{2\kappa}\int^{\Sigma''}_{\Sigma'} d^{3} x
       \sqrt{h} u_{\mu} V^{\mu} 
     + \frac{1}{2\kappa} \int_{\bar{\cal T}}
       d^{3} x \sqrt{- \bar{\gamma}} 
       \bar{n}_{\mu} V^{\mu}
\label{boundaryintegrals}
\end{equation}
as the promised boundary expression for the 
divergence term (\ref{divergenceterm}).

Next, combining the lemma (\ref{lemma}) and its corollary 
(\ref{corollary}) with (\ref{boundaryintegrals}), we write
the divergence term as follows:
\begin{eqnarray}
       (\delta S_{\scriptscriptstyle H})_{\partial {\cal M}} 
& = & 
     - \frac{1}{2\kappa} \int^{B''}_{B'} d^{2}x
       \sqrt{\sigma} \left(n_{\mu} \delta u^{\mu} 
     + \bar{u}_{\mu} \delta \bar{n}^{\mu}\right) 
\nonumber \\
&   & 
     - \frac{1}{2\kappa}\int^{\Sigma''}_{\Sigma'} d^{3} x
       \sqrt{h} (2 \delta K + K^{ij} \delta h_{ij}) 
     + \frac{1}{2\kappa} \int_{\bar{\cal T}}
       d^{3} x \sqrt{- \bar{\gamma}} (
       2\delta \bar{\Theta} + \bar{\Theta}{}^{ij} 
       \delta \bar{\gamma}_{ij} ){\,} .
\label{boundarypluslemma}
\end{eqnarray}
In this expression the total derivative terms in the
boundary element integrals have been expressed as
integrals over the corners $B'$
and $B''$ (again via Stokes' theorem, but now in one lower
dimension). 
Further, for the $\Sigma'$, $\Sigma''$ and $\bar{\cal T}$ integrals
in Eq.~(\ref{boundarypluslemma}) we have chosen to use 
indices adapted to the boundary elements rather than 
general spacetime indices. Now, with the momenta 
(\ref{momenta}) used in 
Eq.~(\ref{boundarypluslemma}), the term
$(\delta S_{\scriptscriptstyle H})_{\partial {\cal M}}$
becomes
\begin{equation}
       (\delta S_{\scriptscriptstyle H})_{\partial {\cal M}} =  
    - \frac{1}{2\kappa} \int^{B''}_{B'} d^{2}x
      \sqrt{\sigma} \left(n_{\mu} \delta u^{\mu} +
      \bar{u}_{\mu} \delta \bar{n}^{\mu}\right) 
     - \int^{\Sigma''}_{\Sigma'} d^{3} x h_{ij} \delta P^{ij}
     - \int_{\bar{\cal T}} d^{3} x \bar{\gamma}_{ij} 
       \delta \bar{\Pi}{}^{ij} {\,} .
\label{momenta_added}
\end{equation}
Our remaining task is to simplify the integrals over $B'$ and
$B''$. To achieve this, recall the boost relations
\begin{eqnarray}
       \bar{u}{}^{\mu} & = & 
       \gamma u^{\mu} + v \gamma n^{\mu}
\label{boostrelations} \eqnum{\ref{boostrelations}a}\\
       \bar{n}{}^{\mu} & = & 
       \gamma n^{\mu} + v \gamma u^{\mu}
\eqnum{\ref{boostrelations}b}
\addtocounter{equation}{1}
\end{eqnarray}
and their inverses, derived in terms of a double foliation of spacetime 
in Appendix A. These can be used to write the integrand of the corner 
integrals as
\begin{equation}
       n_{\mu} \delta u^{\mu} 
     + \bar{u}_{\mu} \delta \bar{n}^{\mu} =
       (\bar{n}{}_{\mu}/\gamma - v u_{\mu}) \delta
       (\bar{u}{}^{\mu}/\gamma - v n^{\mu})
     + (u_{\mu}/\gamma + v \bar{n}{}_{\mu})
       \delta (n^{\mu}/\gamma + v \bar{u}{}^{\mu}) {\,} .
\label{cornerintegrand}
\end{equation}
On the right--hand side of this equation, the terms proportional
to $u_{\mu}\delta n^{\mu}$ vanish. This follows from the 
identity $u_{\mu} \delta n^{\mu} 
 = - n^{\mu} \delta u_{\mu}$ and the fact that $u_{\mu}$ is
hypersurface--orthogonal. [Thus, as seen in identity (ii) after
the Eq.~(\ref{lemma}) $\delta u_{\mu}$ is proportional
to $u_{\mu}$.] Likewise, we have $\bar{n}_{\mu} 
\delta \bar{u}{}^{\mu} = 0$, since $\bar{n}_{\mu}$ is
hypersurface--orthogonal. After a bit of straightforward
algebra, Eq.~(\ref{cornerintegrand}) simplifies to
$n_{\mu} \delta u^{\mu} +
\bar{u}_{\mu} \delta \bar{n}^{\mu} = - 2 \gamma^{2} \delta v$.
Therefore, we may now rewrite Eq.~(\ref{momenta_added}) as
\begin{equation}
(\delta S_{\scriptscriptstyle H})_{\partial {\cal M}}  =  
 \frac{1}{\kappa} \int^{B''}_{B'} d^{2}x
\sqrt{\sigma} \gamma^{2} \delta v 
 - \int^{\Sigma''}_{\Sigma'} d^{3} x h_{ij} \delta P^{ij}
- \int_{\bar{\cal T}}
d^{3} x \bar{\gamma}_{ij} \delta \bar{\Pi}{}^{ij}
{\,} .
\end{equation}
Combination of this result with Eq.~(\ref{deltaS_EH}) and the 
definition $\tanh\theta = v$ of the boost parameter 
yields the desired expression (\ref{mainresult}).

\subsection{Boundary terms and the diffeomorphism 
invariance of the Hilbert action}

The Hilbert action 
(\ref{Einstein_Hilbert_action})
is diffeomorphism invariant. That is, 
the action is unchanged if the variations in the fields 
are given by the Lie derivative along a vector field 
$\xi^\mu$ that is tangent to the boundaries: 
\begin{eqnarray} 
   \delta S_H & = & \frac{1}{2\kappa} \int_{\cal M} 
          d^4x\,{\pounds}_\xi (\sqrt{-g}\Re) 
       = \frac{1}{2\kappa} \int_{\cal M} d^4x 
         \nabla_\mu(\xi^\mu\sqrt{-g}\Re) \cr 
     & = &   - \frac{1}{2\kappa} \int_{\Sigma'}^{\Sigma''} 
        d^3x \sqrt{h}\, u_\mu\xi^\mu \Re
   + \frac{1}{2\kappa} \int_{\barT} d^3x \sqrt{-\bargam}\, 
     \barn_\mu\xi^\mu \Re \cr
   & = & 0 \ .
\end{eqnarray}
Here, we use $\barn_\mu \xi^\mu = 0$ on ${\barT}$, and 
$u_\mu\xi^\mu = 0 $ on $\Sigma'$ 
and $\Sigma''$. These imply $\barn_\mu \xi^\mu = 
u_\mu\xi^\mu = 0 $ on $B'$ and $B''$. 

Since $\delta S_H = 0$ when $\delta$ is given by the Lie derivative,
our main result Eq.~(\ref{mainresult}) implies
\begin{eqnarray} 
    0 =  \delta S_H & = & -\frac{1}{\kappa} \int_{\cal M} 
d^4x \sqrt{-g}\, G^{\mu\nu} 
    \nabla_\mu\xi_\nu + \frac{1}{\kappa} 
\int_{B'}^{B''} d^2x \sqrt{\sigma}\, \gamma^2
   \xi^a\partial_a v  \nonumber\\
   & & -  \int_{\Sigma'}^{\Sigma''} d^3x 
\,h_{ij} {\pounds}_\xi P^{ij} -   \int_{\barT} d^3x
   \, \bargam_{ij} {\pounds}_\xi\barPi^{ij} \, .
\label{anotherresult}
\end{eqnarray} 
Now use the identity 
\begin{eqnarray}
h_{ij}{\pounds}_\xi P^{ij} & = & 
{\pounds}_\xi P - P^{ij}{\pounds}_\xi h_{ij} = D_i(P\xi^i) 
   - 2P^{ij}D_i\xi_j \cr 
  & = & D_i(P\xi^i - 2P^{ij}\xi_j) + 2(D_iP^{ij}) \xi_j \ ,
\end{eqnarray}  
where $\xi^i$ is 
the pullback of $\xi^\mu$ to $\Sigma'$ or $\Sigma''$. Note that 
the $D_i(P\xi^i)$ term 
will vanish when integrated to the corners. Thus, the 
$\Sigma'$ and $\Sigma''$ terms 
in (\ref{anotherresult}) become 
\begin{equation} 
 - \int_{\Sigma'}^{\Sigma''} d^3x \,h_{ij} {\pounds}_\xi P^{ij} 
   = -2\int_{\Sigma'}^{\Sigma''} d^3x \,(D_iP^{ij}) 
\xi_j + 2\int_{B'}^{B''} d^2x \sqrt{\sigma} 
   \,n_i P^{ij}\xi_j/\sqrt{h} \ . 
\end{equation}
Similarly, we find 
\begin{equation} 
 - \int_{\barT} d^3x \,\bargam_{ij} {\pounds}_\xi \barPi^{ij} 
   = - 2\int_{\barT} d^3x \,({\barD}_i\barPi^{ij}) \xi_j - 2\int_{B'}^{B''} d^2x \sqrt{\sigma} 
   \,\baru_i \barPi^{ij}\xi_j/\sqrt{-\bargam} \, ,
\end{equation}
whence Eq.~(\ref{anotherresult}) now becomes 
\begin{eqnarray} 
    0 =  \delta S_H 
   & = & -\frac{1}{\kappa} \int_{\cal M} d^4x \sqrt{-g}(\nabla_\mu G^{\mu\nu}) \xi_\nu 
        + \int_{B'}^{B''} d^2x \sqrt{\sigma}\left( {\bar\jmath}_a\xi^a - \jmath_a\xi^a 
         + \gamma^2\xi^a \partial_a v /\kappa \right) \cr
   & & + \int_{\Sigma'}^{\Sigma''} d^3x \left(\sqrt{h}\, u_\mu G^{\mu\nu}\xi_\nu/\kappa  
      - 2 (D_i P^{ij})\xi_j \right) \cr 
   & & -\int_{\barT} d^3x \left( \sqrt{-\bargam}\, \barn_\mu G^{\mu\nu}\xi_\nu/\kappa 
      + 2(\barD_i\barPi^{ij})\xi_j \right) 
    \ ,\label{yetanotherresult}
\end{eqnarray}
where we have used integration by parts on the 
volume (${\cal M}$) integral term. Also, 
we have used the definitions ${\bar\jmath}_a =  
-2\sigma_{ai}\barPi^{ij}\baru_j/\sqrt{-\bargam}$ 
and $\jmath_a = -2\sigma_{ai}P^{ij}n_j/\sqrt{h}$.

We now use the well--known result that the gravitational field contributions to the 
boundary momentum constraints satisfy 
\begin{eqnarray} 
   {\cal H}_i & = & -2D_jP^j_i = - \sqrt{h} u_\mu G^{\mu\nu} h_{\nu i}/\kappa \ , \\ 
   {\bar{\cal H}}_i & = & -2\barD_j\barPi^j_i = \sqrt{-\bargam}\barn_\mu G^{\mu\nu} 
   \bargam_{\nu i}/\kappa  \ . 
\end{eqnarray} 
Therefore the last two integrals in Eq.~(\ref{yetanotherresult}) vanish. 
Since the result (\ref{yetanotherresult}) must 
hold for all $\xi^\mu$ that are tangent to the boundary, we conclude that 
\begin{eqnarray} 
   & & \nabla_\mu G^{\mu\nu} = 0  \ , \label{Bianchiboost}\eqnum{\ref{Bianchiboost}a}\\
   & & {\bar\jmath}_a = \jmath_a  - \frac{1}{\kappa}\partial_a\theta \ ,
   \eqnum{\ref{Bianchiboost}b}
   \addtocounter{equation}{1}
\end{eqnarray} 
where $\theta$ is the velocity parameter, $v = \tanh(\theta)$. Equation 
(\ref{Bianchiboost}a) is, of course, the contracted Bianchi identity. Equation 
(\ref{Bianchiboost}b) is an identity as well. In fact, as we will see in the 
next section, ${\bar\jmath}_a$ and 
$\jmath_a$ are the tangential momentum densities for the barred and unbarred observers, 
and the identity (\ref{Bianchiboost}b) expresses the boost relationship between 
these quantities. 

Note that this analysis can be applied to the Trace-K action as well. Indeed, any 
action that is diffeomorphism invariant and differs from the Hilbert action by boundary 
terms can be used. The reason is that the Lie variation of a boundary term will 
always integrate to the corners, and then vanish since 
$\xi^\mu$ is tangent to the corner.


\section{Quasilocal Stress-Energy-Momentum and Boost Relations} 
\subsection{Quasilocal quantities}
Using our main result (\ref{mainresult}) for the variation of the Hilbert
action, one can easily show that the variation of the Trace-K action 
(\ref{TraceKaction}) has
the following boundary terms:
\be
(\delta S)_{\partial\M} =
\int^{\Sigma''}_{\Sigma'} d^{3} x P^{ij} \delta h_{ij}
+ \int_{\bar{\cal T}} d^{3} x \bar{\Pi}{}^{ij} \delta \bar{\gamma}{}_{ij}
- \frac{1}{\kappa} \int^{B''}_{B'} d^{2}x \theta \delta \sqrt{\sigma} 
\ .
\label{varyTraceK}
\ee
Notice that the Trace-K action features solely fixation of the induced metric
on the boundary $\partial\M$.
We now wish to express the $\barT$ boundary term in $\delta S$ in terms 
of the geometry of the $\Sigma$ slices. Start with the $(\delta S)_{\barT}$
contribution to the variation, that is
\be 
\int_{\barT} d^3x \, \barPi^{ij} \delta\bargam_{ij} = -\frac{1}{2\kappa} 
\int_{\barT} d^3x \sqrt{-\bargam}  (\barTheta \bargam^{ij} - \barTheta^{ij} ) 
\delta \bargam_{ij} \ .
\ee 
With an ADM splitting of the side boundary metric $\bargam_{ij}$ into 
a lapse function $\barN$, a shift vector $\barV^a$, and a spatial metric 
$\sigma_{ab}$ (see Appendix A), we find 
\be 
\delta \bargam_{ij} = -\frac{2}{\barN}  \baru_i \baru_j \delta\barN 
- \frac{2}{\barN} \sigma_{a{\sss (}i} \baru_{j{\sss )}} \delta \barV^a 
+ \sigma^a_{{\sss (}i} \sigma^b_{j{\sss )}} \delta\sigma_{ab} \ . 
\ee 
With this splitting of the $\barT$ metric we then obtain 
\bea 
\int_{\barT} d^3x \, \barPi^{ij} \delta\bargam_{ij} = -\frac{1}{\kappa} 
\int_{\barT} d^3x \sqrt{\sigma} \biggl\{ 
(\barTheta + && \baru_i\barTheta^{ij}\baru_j)\delta\barN 
+ (\sigma_{ai}\barTheta^{ij}\baru_j)\delta\barV^a \nonumber\\
&& + \frac{\barN}{2} (\barTheta \sigma^{ab} - \sigma^a_i\barTheta^{ij} \sigma^b_j) 
\delta\sigma_{ab} \biggr\} \ . 
\eea 
Now, in order to achieve our goal of expressing $(\delta S)_{\barT}$
in terms of $\Sigma$ geometry, we must find a ``splitting" of the $\barT$
extrinsic curvature tensor $\barTheta_{ij}$. The desired expression
\be 
\barTheta_{\mu\nu} = \gamma k_{\mu\nu} + \gamma v \ell_{\mu\nu} + 
(\barn \cdot \bara) \baru_\mu \baru_\nu  + 2\sigma^\alpha_{{\sss (}\mu} 
\baru_{\nu{\sss )}} (K_{\alpha\beta} n^\beta - \nabla_\alpha \theta) \ ,
\label{barTsplit}
\ee 
is derived in Appendix B. In this expression, we have used the definitions 
\bea 
k_{\mu\nu} 
&=& -\sigma^\alpha_\mu \sigma^\beta_\nu \nabla_\alpha n_\beta \ ,
\label{kandl}\eqnum{\ref{kandl}a}\\
\ell_{\mu\nu} 
&=& -\sigma^\alpha_\mu \sigma^\beta_\nu \nabla_\alpha u_\beta \ .
\eqnum{\ref{kandl}b} 
\addtocounter{equation}{1}
\eea 
The unit normals $u^\mu$, $n^\mu$ associated with the hypersurfaces $\Sigma$  are 
related to the unit normals $\baru^\mu$, $\barn^\mu$ associated with $\barT$ 
as in Eqs.~(\ref{boostrelations}). Again, our conventions are that barred 
observers are comoving with the boundary $\barT$ while the unbarred ones are
at rest in the $\Sigma$ hypersurfaces. Also in Eq.~(\ref{barTsplit}), 
$\bara^\mu= \baru^\nu
\nabla_\nu \baru^\mu$ denotes the acceleration  of the barred observers, 
and $K_{\alpha\beta}$ denotes the extrinsic curvature  of the $\Sigma$ slices. 
Putting these results together, we have 
\bea 
\int_{\barT} d^3x \, \barPi^{ij} \delta\bargam_{ij} = -\frac{1}{\kappa} 
\int_{\barT} d^3x \sqrt{\sigma} \biggl\{ &&
[\gamma k + \gamma v \ell ]\delta\barN 
- [\sigma_a^i K_{ij} n^j - \partial_a\theta]\delta\barV^a \nonumber\\
&&- \frac{\barN}{2} \Bigl[ \gamma(k^{ab} - k\sigma^{ab}) + 
\barn\cdot\bara\, \sigma^{ab} + \gamma v (\ell^{ab} - \ell\sigma^{ab}) \Bigr] 
\delta\sigma_{ab}\biggr\}  \label{SigmabarT} 
\eea
for the $\barT$ term in the variation of the action. 

The contribution to $\delta S$ from the top and bottom 
caps ($\Sigma''$ and $\Sigma'$) is 
\be 
\int^{\Sigma''}_{\Sigma'} d^3x \, P^{ij} \delta h_{ij} 
= \frac{1}{2\kappa} \int^{\Sigma''}_{\Sigma'} d^3x \sqrt{h} 
(K h^{ij} - K^{ij})\delta h_{ij} \ . 
\ee 
The induced metric $h_{ij}$ can be split into a ``radial" lapse function 
$M$, shift vector $W^a$, and slice metric $\sigma_{ab}$ (see Appendix A). 
The variation in $h_{ij}$ is then 
\be 
\delta h_{ij} = \frac{2}{M} n_i n_j \delta M 
+ \frac{2}{M} \sigma_{a{\sss (}i} n_{j{\sss )}} \delta W^a 
+ \sigma^a_{{\sss (}i} \sigma^b_{j{\sss )}} \delta\sigma_{ab} \ , 
\ee 
from which we obtain 
\bea 
\int^{\Sigma''}_{\Sigma'} d^3x \, P^{ij} \delta h_{ij} 
= \frac{1}{\kappa} \int^{\Sigma''}_{\Sigma'} d^3x \sqrt{\sigma} \biggl\{ 
 (K - && n_i K^{ij} n_j)\delta M 
- (\sigma_{ai} K^{ij} n_j)\delta W^a \nonumber\\ 
&& + \frac{M}{2}(K\sigma^{ab} - \sigma^a_i K^{ij}\sigma^b_j) 
\delta\sigma_{ab} \biggr\} 
\eea
Now use the splitting 
\be 
K_{\mu\nu} = \ell_{\mu\nu} + (u\cdot b) n_\mu n_\nu  + 2\sigma^\alpha_{{\sss (}\mu} 
n_{\nu{\sss )}} K_{\alpha\beta} n^\beta  \label{Ksplit}
\ee 
from Appendix B, where $b^\mu = n^\nu \nabla_\nu n^\mu$. These 
results yield 
\be
\int^{\Sigma''}_{\Sigma'} d^3x \, P^{ij} \delta h_{ij} 
= \frac{1}{\kappa} \int^{\barSigma''}_{\barSigma'} d^3x \sqrt{\sigma} \biggl\{ 
\ell\delta M 
- \sigma^i_a K_{ij} n^j \delta W^a 
 - \frac{M}{2} \Bigl[(\ell^{ab} - \ell\sigma^{ab}) 
- u\cdot b\, \sigma^{ab}\Bigr]
\delta\sigma_{ab} \biggr\} \label{caps} 
\ee
for the top and bottom--cap terms in the variation of the action. 

The result (\ref{SigmabarT}) allows us to define the  quasilocal densities 
associated with the two--surfaces $B$ as seen by the 
``barred" observers: 
\bea
\kappa\barepsi &\equiv & -\frac{\kappa}{\sqrt{\sigma}}\frac{\delta
S\bigr|_{\barT}}{\delta\barN}  = \gamma k + \gamma v \ell \ ,
\label{bardensities}\eqnum{\ref{bardensities}a} \\
\kappa\barj_a &\equiv &  \frac{\kappa}{\sqrt{\sigma}}\frac{\delta
S\bigr|_{\barT}}{\delta\barV^a}  
= \sigma^i_a K_{ij} n^j - \partial_a\theta 
\ ,\eqnum{\ref{bardensities}b} \\
\kappa\bars^{ab} &\equiv& \frac{2\kappa}{\barN\sqrt{\sigma}} 
\frac{\delta S\bigr|_{\barT}}{\delta\sigma_{ab}} 
= \gamma(k^{ab} - k\sigma^{ab}) + 
\barn\cdot\bara\, \sigma^{ab} + \gamma v (\ell^{ab} - \ell\sigma^{ab}) 
\ .\eqnum{\ref{bardensities}c}
\addtocounter{equation}{1}
\eea 
These are the quasilocal energy density, tangential momentum density, and 
spatial stress, respectively. The notation $\delta S\bigr|_{\barT}$ refers 
to a {\sc hj} variation of the Trace-K action $S$, with respect to the $\barT$ 
metric components $\barN$, $\barV^a$, and $\sigma_{ab}$. These definitions 
hold for each leaf of the $\barT$--foliation $B$, but our attention will be 
focused primarily on the corner $B''$. 
Likewise, the result (\ref{caps}) allows us to define quasilocal densities
as seen by the ``unbarred'' observers:
\bea
\kappa \jperp &\equiv & -\frac{\kappa}{\sqrt{\sigma}}\frac{\delta
S\bigr|_{\Sigma''}}{\delta M}  = - \sigma^{ij}K_{ij} = 
- \ell\ ,\label{unbardensities}
\eqnum{\ref{unbardensities}a} \\
\kappa \jmath_a &\equiv & -\frac{\kappa}{\sqrt{\sigma}}\frac{\delta 
S\bigr|_{\Sigma''}}{\delta W^a}
= \sigma^i_a K_{ij} n^j \eqnum{\ref{unbardensities}b} \\
\kappa t^{ab} &\equiv& \frac{2\kappa}{M\sqrt{\sigma}}
\frac{\delta S\bigr|_{\Sigma''}}{\delta\sigma_{ab}}
= K\sigma^{ab} - \sigma^a_i K^{ij}\sigma^b_j = 
- (\ell^{ab} - \ell\sigma^{ab})
+ u\cdot b\, \sigma^{ab} \ .\eqnum{\ref{unbardensities}c}
\addtocounter{equation}{1}
\eea
These are the quasilocal normal momentum
density, tangential momentum density, and temporal stress, respectively.
The notation
$\delta S\bigr|_{\Sigma''}$ refers to a {\sc hj} variation of the Trace-K 
action $S$ with respect to the $\Sigma''$ metric components $M$, $W^a$, and 
$\sigma_{ab}$. These definitions hold for each slice of the ``radial" foliation 
of $\Sigma''$, but again we focus attention on the corner $B''$. 

Clearly the definitions (\ref{bardensities}) and (\ref{unbardensities}) are 
applicable to any closed two--dimensional surface $B$ embedded in a spacetime 
that satisfies the Einstein equations---we simply arrange to have the top corner $B''$ 
of the manifold $\M$ coincide with the given surface $B$ and apply the definitions. 
The surface $B$ can be pierced by various fleets of observers,  for example, barred 
and unbarred observers. Different observers who are  boosted relative to one another 
will  see different quasilocal densities for the same surface
$B$. With this in mind and to put the set (\ref{unbardensities}) on an equal 
footing with the set (\ref{bardensities}), we define additional barred densities
\bea
\kappa\barjperp &\equiv & -\sigma^{ij} \barK_{ij} = 
-\gamma \ell - \gamma v k \, ,
\eqnum{\ref{bardensities}d} \\
\kappa\barj_a &\equiv & \sigma^i_a \barK_{ij} \barn^j =
\sigma^i_a K_{ij} n^j - \partial_a\theta \, ,
\eqnum{\ref{bardensities}e} \\
\kappa\bart^{ab} &\equiv& 
\barK\sigma^{ab} - \sigma^a_i \barK^{ij}\sigma^b_j
= -\gamma(\ell^{ab} - \ell\sigma^{ab})
+ \baru\cdot\barb\, \sigma^{ab} - \gamma v(k^{ab}- k\sigma^{ab})
\ .\eqnum{\ref{bardensities}f}
\eea
Note that these expressions are defined in terms of a slice $\barSigma$ [with
intrinsic and extrinsic geometry $(\barh_{ij}, \barK^{ij})$] which meets the
$\barT$ boundary orthogonally. Observers comoving with $\barT$ are at rest with 
respect to $\barSigma$.${}^{10}$
It is not difficult to see that in 
terms of the $(\barh_{ij}, \barK^{ij})$ geometry, $\barjperp$, $\barj_a$, 
and $\bart^{ab}$ have exactly the same forms as do 
$\jperp$, $\jmath_a$, and $t^{ab}$ in terms of $(h_{ij}, K^{ij})$
geometry.
In the rightmost expressions we have expressed $\barjperp$, $\barj_a$,
and $\bart^{ab}$ in terms of $\Sigma$ geometry, using a ``splitting''
similar to the one given in Eq.~(\ref{barTsplit}) but this time expressing
the spacetime representation $\barK_{\mu\nu}$ of the $\barSigma$ extrinsic 
curvature tensor in terms of $\Sigma$ geometry. The relevant splitting is
found in Appendix B. Finally, note that expressions (\ref{bardensities}b)
and (\ref{bardensities}e) agree.

In Eqs.~(\ref{bardensities}), 
the quasilocal densities for the barred observers are expressed 
in terms of the geometry  and foliation defined by the unbarred observers. 
Alternatively, those densities for the barred observers can be expressed 
in terms of the geometry and foliation defined by the barred observers themselves. 
This is achieved by keeping the boundary $\partial\M$ fixed in a neighborhood of $B''$, 
and tilting the $\Sigma$ slices until the unbarred observers coincide with the barred
observers. In other words $\Sigma$ slices become $\barSigma$ slices.  
The boost velocity $v$ then vanishes, and Eqs.~(\ref{bardensities}) become 
\bea 
\kappa\barepsi &=& \bark \ ,
\label{newbar}\eqnum{\ref{newbar}a}\\
\kappa\barjperp &=& -\barell \ ,\eqnum{\ref{newbar}b}\\
\kappa\barj_a &=& \sigma_a^i \barK_{ij} \barn^j \ ,\eqnum{\ref{newbar}c}\\
\kappa \bars^{ab} &=& \bark^{ab} - \bark\sigma^{ab} 
+ (\barn\cdot \bara)\sigma^{ab} \ ,\eqnum{\ref{newbar}d}\\
\kappa \bart^{ab} &=& -\barell^{ab} + \barell\sigma^{ab} 
+ (\baru\cdot \barb)\sigma^{ab}  \ .\eqnum{\ref{newbar}e}
\addtocounter{equation}{1}
\eea 
Here, the barred quantities $\bark^{ab}$, $\barK_{ij}$, {\it etc.\/} refer 
to the surface $B''$ embedded in the top cap $\barSigma''$. 
The results (\ref{newbar}) extend the definitions given in the original QLE 
paper \cite{BY}${}^{11}$ to 
include the normal momentum density $\barjperp$ and the temporal stress 
tensor $\bart^{ab}$.   Of course, we can view the limit $v\to 0$ of 
Eqs.~(\ref{bardensities}) in another way: consider the unbarred observers 
($\Sigma$ slices) as unchanged, and the boundary $\partial\M$ at the 
corner $B''$ as ``unboosted" until the barred 
observers coincide with the unbarred observers. Then we obtain the relationships 
(\ref{newbar}), but without the bars. That is, we find that the energy surface density 
for the unbarred observers is $\kappa\epsi = k$, with similar expressions for 
the momentum densities and stress tensors. 

Before continuing with the main line of reasoning, let us discuss the 
physical significance of the normal and tangential momentum densities. The normal momentum
density can be  written as $\kappa\jperp  = \sigma^{\mu\nu}\nabla_\mu u_\nu = 
-\sigma^{\mu\nu} K_{\mu\nu} = n^i n^j (K_{ij} - K h_{ij})$, and the 
tangential momentum density can be written as 
$\kappa \jmath_a = \sigma_a^i n^j (K_{ij} - K h_{ij})$. These quantities are the 
normal and tangential components of the (total) momentum surface density 
$\kappa\jmath^i = n_j(K^{ij} - Kh^{ij})$, which can be written in terms of 
the gravitational momentum as $\jmath^i = -2 P^{ij} n_j/\sqrt{h}$. We now remark  
that the analysis presented in this paper can be easily generalized to 
include matter fields. For the case of nonderivative coupling (in 
which the matter action does not contain derivatives of the metric) 
the basic definitions (\ref{newbar}) are unchanged. By including 
matter fields in the definition of the system we find that 
$\jmath^i = \jperp n^i + \jmath_a \sigma^{ai}$ is
related to the  matter momentum in the following way. Consider the momentum 
constraint 
\be 
0 = -2 D_i P^{ij} + \sqrt{h} u_\mu T^{\mu j} 
\ee 
for the hypersurfaces $\Sigma$, where $- u_\mu T^{\mu j}$ 
is the proper matter momentum density in the $j$th
direction. Assume that there exists a Killing vector field $\xi^i$ 
on space $\Sigma$. It is straightforward to show that the total 
matter momentum along $\xi^i$ is 
\be 
-\int_\Sigma d^3x \sqrt{h} u_\mu T^{\mu i} \xi_i  =  -2 
\int_\Sigma d^3x D_i P^{ij}  \xi_j 
=  \int_B d^2x \sqrt{\sigma} \jmath_i \xi^i \ , 
\ee 
where $B = \partial\Sigma$. 
This shows that $\jmath^i$ represents a surface density for the matter momentum. 

\subsection{Boost relations}
We now return to Eqs.~(\ref{bardensities}), expressing the quasilocal densities 
for the barred observers. In terms of the unboosted quasilocal densities [the unbarred versions of 
Eqs.~(\ref{newbar})], Eqs.~(\ref{bardensities}a,d) become 
\bea 
\barepsi &=& \gamma \epsi - \gamma v \jperp \ ,
\label{epjboosts}\eqnum{\ref{epjboosts}a}\\
\barjperp &=& \gamma \jperp - \gamma v \epsi \ ,
\eqnum{\ref{epjboosts}b}
\addtocounter{equation}{1}
\eea 
relating the quasilocal energy density and the 
normal momentum density for barred and unbarred observers. We also obtain 
\be
\barj_a = \jmath_a - \partial_a\theta/\kappa \ ,\label{tanjboost}
\ee 
for the tangential momentum density. 
Finally, we have 
\bea
\Bigl[ \bars^{ab} - (\barn\cdot\bara)\sigma^{ab}/\kappa \Bigr] 
&=& \gamma \Bigl[ s^{ab} - (n\cdot a)\sigma^{ab}/\kappa \Bigr] 
-\gamma v \Bigl[ t^{ab} - (u\cdot b)\sigma^{ab}/\kappa \Bigr] \ ,
\label{stressboost}\eqnum{\ref{stressboost}a}\\
\Bigl[ \bart^{ab} - (\baru\cdot\barb)\sigma^{ab}/\kappa \Bigr] 
&=& \gamma \Bigl[ t^{ab} - (u\cdot b)\sigma^{ab}/\kappa \Bigr] 
-\gamma v \Bigl[ s^{ab} - (n\cdot a)\sigma^{ab}/\kappa \Bigr] \ .
\eqnum{\ref{stressboost}b}
\addtocounter{equation}{1}
\eea 
for the boost relation between the spatial and temporal stress tensors. 
This later relation can be rewritten using the results 
\bea 
\barn \cdot \bara 
&=& \gamma n\cdot a - \gamma v u\cdot b + \baru\cdot\nabla\theta \ ,
\label{accboost}\eqnum{\ref{accboost}a}\\
\baru \cdot \barb 
&=& \gamma u\cdot b - \gamma v n\cdot a - \barn\cdot\nabla\theta \ ,
\eqnum{\ref{accboost}b}
\addtocounter{equation}{1}
\eea 
from Appendix B. We thus obtain 
\bea 
\bars^{ab} &=& \gamma s^{ab} - \gamma v t^{ab} 
+ (\baru\cdot\nabla\theta/\kappa)\sigma^{ab} \ ,
\label{stressboost2}\eqnum{\ref{stressboost2}a}\\
\bart^{ab} &=& \gamma t^{ab} - \gamma v s^{ab} 
- (\barn\cdot\nabla\theta\kappa)\sigma^{ab} \ ,
\eqnum{\ref{stressboost2}b}
\addtocounter{equation}{1}
\eea
for the boost relation satisfied by $s^{ab}$ and $t^{ab}$. 
Finally, let us define the spatial shear $\eta^{ab} = s^{ab} - s\sigma^{ab}/2$ as the
trace--free  part of the spatial stress $s^{ab}$, and the temporal shear $\zeta^{ab} = t^{ab} 
- t\sigma^{ab}/2$ as the trace--free part of the temporal stress $t^{ab}$. From 
the unbarred version of Eq.~(\ref{newbar}), we have 
\bea 
\kappa\eta^{ab} &=& k^{ab} - k\sigma^{ab}/2 \ ,
\label{tracefree}\eqnum{\ref{tracefree}a}\\
\kappa\zeta^{ab} &=& -\ell^{ab} + \ell\sigma^{ab}/2 \ .
\eqnum{\ref{tracefree}b}
\addtocounter{equation}{1}
\eea
The results (\ref{stressboost}), or equivalently (\ref{stressboost2}), yield 
\bea
\bareta^{ab} &=& \gamma \eta^{ab} - \gamma v \zeta^{ab} \ ,
\label{tfboost}\eqnum{\ref{tfboost}a}\\
\barzeta^{ab} &=& \gamma\zeta^{ab} - \gamma v \eta^{ab} \ ,
\eqnum{\ref{tfboost}b} 
\addtocounter{equation}{1}
\eea
for the boost relation between $\eta^{ab}$ and $\zeta^{ab}$. 

\subsection{Boost Invariants}

The results (\ref{epjboosts}a,b) show that the energy surface 
density and normal momentum density behave under local boosts 
like the time and space components of an energy--momentum vector, 
namely, 
$\epsi u^\mu + \jperp n^\mu = 
\barepsi\baru^\mu + \barjperp \barn^\mu$. 
Clearly, the squared length of the vector 
$\epsi u^\mu + \jperp n^\mu$, defined by 
\be 
M^2/\kappa^2 = \varepsilon^2 - \jmath_\vdash^2 \ ,\label{invariant1}
\ee 
is invariant under boosts. We do not claim that $M^2$ is in all cases
positive. However, if it is, then $M/\kappa$ 
(defined via the negative
square root\cite{Lau3}) is equal to $\bar{\varepsilon}$ for a fleet of  
observers $\bar{u}^\mu$ who pass through $B$ in such a way that 
$\bar{\jperp} = -\bar{\ell}/\kappa = 0$;  that is, such that $B$ 
is a maximal slice of $\bar{\T}$ (if such a slice exists). This defines 
locally, at each point of $B$, a rest 
frame for the system. Moreover, the parameter 
associated with the local 
boost between an arbitrary frame and the rest 
frame can be computed
via\cite{Lau3}
\be
\theta = {\textstyle \frac{1}{2}} \log\left[ 
\frac{\varepsilon + \jperp}{\varepsilon - \jperp}\right] \, .
\label{boostparam}
\ee
Indeed, using the relations inverse to those given in 
Eqs.~(\ref{epjboosts}a,b) along with the rest--frame 
condition $\bar{\jperp} = 0$, we find that
\be
\theta = {\textstyle \frac{1}{2}} \log\left[
\frac{1+v}{1-v}\right] \, ,
\ee
which is immediately recognized as the logarithmic 
representation of $\tanh^{-1}(v)$. Note that 
Eq.~(\ref{boostparam}) demonstrates that the two-surface 
data $\{\varepsilon,\jperp\}$ encodes the rest frame 
direction. This fact features prominently in Kucha\v{r}'s 
examination of the geometrodynamics of
Schwarzschild black-holes.\cite{Kuchar,Lau3}

Equation (\ref{tanjboost}) expresses the change in the tangential 
momentum surface density $\jmath_a$ under a boost. 
Evidently the curl of $\jmath_a$, 
\be 
F_{ab} = \partial_a \jmath_b - \partial_b \jmath_a \ , 
\label{Foftanj}
\ee 
is invariant under boosts. 
Also note that $\jmath_a$ itself is invariant under boosts that are 
constant on $B$. 

We now turn to the spatial and temporal shear. The boost 
relations (\ref{tfboost}a,b) show that the shear tensors 
transform like the components of a (two--dimensional, 
traceless, symmetric matrix valued) 
spacetime vector ${\tilde H}^{ab\mu} = \kappa(\eta^{ab} u^\mu + 
\zeta^{ab} n^\mu)$. The shear tensors can be combined to form 
boost invariants, such as 
\bea 
(M_1)^2 & = & {\tilde H}^{ab\mu} {\tilde H}_{ab\mu}
= \kappa^{2}( \eta^{ab}\eta_{ab} - \zeta^{ab}\zeta_{ab}) \, ,
\eqnum{\ref{invariants123}a}\\ 
(M_2)^4 & = & {\tilde H}^{ab\mu} {{\tilde H}^{cd}}_\mu 
{{\tilde H}_{ab}}^\nu {\tilde H}_{cd\nu} - 
({\tilde H}^{ab\mu} {\tilde H}_{ab\mu})^2  =  
2\kappa^{4}
\eta^{ab} ( \eta_{ab} \zeta_{cd} - \zeta_{ab} \eta_{cd} )
\zeta^{cd} \, ,
\label{invariants123} \eqnum{\ref{invariants123}b}\\
(M_3)^2 & = & \epsilon^{ab\mu\nu}{\tilde H}^{c}{}_{a\mu} 
{\tilde H}_{bc\nu}
= 2\kappa^{2}\epsilon^{ab}\eta^{c}{}_{a}\zeta_{bc} \, ,
\eqnum{\ref{invariants123}c}
\addtocounter{equation}{1}
\eea 
where $\epsilon_{ab} = u^{\mu} n^{\nu} \epsilon_{\mu\nu a b}$ 
is the alternating tensor on $B$. Again, we do not claim 
that $(M_1)^2$, $(M_2)^4$, and $(M_3)^2$ are positive.  

With the invariants $(M_1)^2$, $(M_2)^4$, and $(M_3)^2$ we can
build several different mass definitions that have appeared in the literature.\cite{Lau2}
In terms of the expansions $\rho$ and $\mu$ of the null normals
$(u^{\mu} \pm n^{\mu})/\sqrt{2}$ to $B$ (spin coefficients in
the Newmann-Penrose formalism\cite{Tod}), we have that
$M^{2} = 8\mu\rho$. Hence the familiar Hawking
mass\cite{Hawking} may be expressed as
\be
M_{\rm Hawking} = \frac{1}{\kappa}\sqrt{\frac{A}{16\pi}} \int_{B}
d^{2}x \sqrt{\sigma}\left[-(M^{2}/2) 
+ {\cal R}\right]\, ,
\ee
where $\cal R$ is the Ricci scalar and 
$A$ the area of $B$. The prefactor in front of 
the integral ensures that the overall expression has 
units of inverse length (i.~e.~energy in geometrical units). 

The following is a geometric identity relating the $\cal M$
Riemann tensor $\Re_{\alpha\beta\mu\nu}$
with the two--surface data of $B$:\cite{Szabados0}
\be
\sigma^{\mu\sigma}\sigma^{\lambda\kappa} 
\Re_{\mu\lambda\sigma\kappa} = k^{ab} k_{ab} - k^2 - \ell_{ab}
\ell^{ab} + \ell^2 + {\cal R} = 
(M_1)^2 - (M^2/2) + {\cal R}\, ,
\ee
where here the $B$ two--metric 
$\sigma_{\mu\nu} = g_{\mu\nu} - n_{\mu} n_{\nu}
+ u_{\mu} u_{\nu}$ serves as a projection operator.
Hayward's quasilocal mass\cite{SHayward} is
\be
M_{\rm Hayward} = \frac{1}{\kappa}\sqrt{\frac{A}{16\pi}} \int_{B}        
d^{2}x \left(\sqrt{\sigma}\sigma^{\mu\sigma}\sigma^{\lambda\kappa}
\Re_{\mu\lambda\sigma\kappa} + \sigma_{\mu\nu} [u,n]^{\mu} 
[u,n]^{\nu}\right) \, ,
\label{Haywardmass}
\ee
where $[u,n]^{\mu}$ is the vector-field commutator between the
$B$ normals. One may verify that the last term in Hayward's
mass is boost invariant, although it would not seem expressible
solely in terms of the two--surface data of $B$. Striking this
term from the integrand one obtains an energy expression which
has proved useful in asymptotic investigations.\cite{AshHan}

\subsection{Second Fundamental Form of $B$ in $\cal M$}

The second fundamental form (extrinsic curvature) for a spacelike 
two--dimensional surface $B$ embedded in four--dimensional spacetime 
$\M$ is defined by 
\be 
{H_{\alpha\beta}}^\mu = \sigma_\alpha^\gamma \sigma_\beta^\delta 
\nabla_\gamma \sigma_\delta^\mu \ . 
\ee 
Here, as always, $\sigma_{\alpha\beta}$ is the induced 
metric on $B$ and $\nabla$ is the covariant derivative 
in $\M$. With the representation $\sigma_{\alpha\beta} = 
g_{\alpha\beta} + u_\alpha u_\beta - n_\alpha n_\beta$, 
the second fundamental form becomes 
\be 
{H_{\alpha\beta}}^\mu = \sigma_\alpha^\gamma \sigma_\beta^\delta  
(u^\mu \nabla_\gamma u_\delta  - n^\mu \nabla_\gamma n_\delta) 
\ .
\ee 
From this result it follows that $u_\mu {H_{\alpha\beta}}^\mu = 
\ell_{\alpha\beta}$ is the extrinsic curvature of $B$ as 
a surface embedded in $\T$, where $\T$ is the three--dimensional 
spacetime orthogonal to $n^\mu$. It also follows that 
$n_\mu {H_{\alpha\beta}}^\mu = k_{\alpha\beta}$ is the extrinsic 
curvature of $B$ as a surface embedded in $\Sigma$, where $\Sigma$ 
is the three--dimensional space orthogonal to $u^\mu$. Thus, the 
second fundamental form of $B$ is 
\be 
{H_{\alpha\beta}}^\mu = k_{\alpha\beta} n^\mu - \ell_{\alpha\beta} 
u^\mu \ .
\ee 
In another basis $\baru^\mu$, $\barn^\mu$ for the spacetime 
orthogonal to $B$, the second fundamental form becomes 
\be 
{H_{\alpha\beta}}^\mu = \bark_{\alpha\beta} \barn^\mu - 
\barell_{\alpha\beta} \baru^\mu \ ,
\ee 
where $\barell_{\alpha\beta}$ is the extrinsic curvature for $B$ 
embedded in $\barT$ (which is orthogonal to $\barn^\mu$), and 
$\bark_{\alpha\beta}$ is the extrinsic curvature for $B$ 
embedded in $\barSigma$ (which is orthogonal to $\baru^\mu$). 
By using the boost relations (\ref{barrelations}a,b)
we find that the $\barn^\mu$ and 
$\baru^\mu$ components of ${H_{\alpha\beta}}^\mu$ are 
\bea 
\bark_{\alpha\beta} & = & \gamma k_{\alpha\beta} + 
\gamma v \ell_{\alpha\beta} \ ,
\label{fundformboost}\eqnum{\ref{fundformboost}a} \\
\barell_{\alpha\beta} & = & \gamma \ell_{\alpha\beta} + 
\gamma v k_{\alpha\beta} \ .\eqnum{\ref{fundformboost}b} 
\addtocounter{equation}{1}
\eea 
Recall that the energy and normal momentum densities are defined by 
$\epsi = k/\kappa$ and 
$\jperp = -\ell/\kappa$, respectively. We therefore 
see that the traces of Eqs.~(\ref{fundformboost}a,b) yield 
the boost relations (\ref{epjboosts}a,b). 
Also recall that the shear tensors are defined by $\kappa\eta^{ab} = 
k^{ab} - k\sigma^{ab}/2$ and $\kappa\zeta^{ab} 
= -\ell^{ab} + \ell\sigma^{ab}/2$. The trace--free parts of 
Eqs.~(\ref{fundformboost}a,b) are then seen to yield 
boost relations (\ref{tfboost}a,b).

The boost invariants among $\epsi$, $\jperp$, 
$\eta^{ab}$, and $\zeta^{ab}$ 
are scalars constructed
from the second  fundamental form 
${H_{\alpha\beta}}^\mu$. Note that contraction 
between upper and lower indices gives zero, since 
$u^\mu \sigma_{\mu\nu} = 0$ and $n^\mu \sigma_{\mu\nu}=0$. 
Thus, nontrivial scalars are formed from 
\bea
{H_{\alpha\beta}}^\mu {H_{\gamma\delta}}_\mu & &
\label{lasteq} \eqnum{\ref{lasteq}a} \\
\epsilon_{\lambda\rho\mu\nu} {H_{\alpha\beta}}^\mu
{H_{\gamma\delta}}^\nu & &
\eqnum{\ref{lasteq}b}
\addtocounter{equation}{1}
\eea 
by contracting the free indices in various ways. For example, 
the invariant $M^2$ is obtained from Eq.~(\ref{lasteq}a) by
contraction 
with $\sigma^{\alpha\beta}\sigma^{\gamma\delta}$, while the invariant 
$(M_1)^2 + M^2/2$ is obtained from Eq.~(\ref{lasteq}a) 
by contraction with 
$\sigma^{\alpha\gamma}\sigma^{\beta\delta}$. The invariant 
$(M_2)^2$ is defined by 
\be (M_2)^4 + (M_1)^4 + M^4/4 = 
H^{\alpha\beta\mu}H^{\gamma\delta}{}_\mu 
H_{\gamma\delta}{}^\nu H_{\sigma\rho\nu}
(\delta^\sigma_\alpha \delta^\rho_\beta 
- \sigma_{\alpha\beta} \sigma^{\sigma\rho}) \, .
\ee
Finally note that one may obtain $(M_{3})^2$ via contraction
of Eq.~(\ref{lasteq}b) with $\sigma^{\lambda\beta} 
\sigma^{\rho\delta} \sigma^{\alpha\gamma}$.

\section{Canonical Theory} 
In this section we consider the Hamiltonian formalism as it pertains
to our bounded spacetime region $\M$. We begin by casting the 
Trace-K action (\ref{TraceKaction}) into canonical form and
examining the canonical variational principle. Next, we compute the
variation of the gravitational Hamiltonian, using a lower--dimensional
version of the lemma proved in Section II.A which was
instrumental in computing the variation of the Hilbert action 
(\ref{Einstein_Hilbert_action}). Finally, we show that the boost
relations (\ref{epjboosts}) and (\ref{tanjboost}) can be obtained from 
the canonical theory.

\subsection{Canonical action principle}
To write the Trace-K action (\ref{TraceKaction}) in terms of the 
canonical variables $(h_{ij}, P^{ij})$, we first insert  the 
space--time split of the spacetime curvature scalar $\R$,\cite{York,BY} 
\be 
\R = R + K_{\mu\nu}K^{\mu\nu} - K^2 - 2\nabla_\mu(K u^\mu + a^\mu) 
\ ,\label{spacetimeR} 
\ee
and the space--time split (\ref{SigmabarT}) of the $\barT$ 
extrinsic curvature $\barTheta_{\mu\nu}$ into the action 
(\ref{TraceKaction}), thereby finding
\be 
S = \frac{1}{2\kappa} \int_\M d^4x \, N\sqrt{h} \bigl(R + 
K_{\mu\nu}K^{\mu\nu} - K^2\bigr)  - \frac{1}{\kappa} \int_{\barT} d^3x\, 
\barN\sqrt{\sigma} \bigl(\gamma k -  \baru\cdot\nabla\theta \bigr) 
-\frac{1}{\kappa}\int^{B''}_{B''}d^{2}x \sqrt{\sigma}\theta\ .
\label{halfway}\ee 
In deriving this expression, we have used Stokes'
theorem, the result $K = h^{\mu\nu}K_{\mu\nu} 
= -(\sigma^{\mu\nu} + n^\mu n^\nu)\nabla_\mu u_\nu = \ell 
+ u\cdot b$ along with $\barn \cdot a = \gamma n\cdot a$, 
the standard identity $\sqrt{-g} = N\sqrt{h}$, and Eq.~(\ref{accboost}a). 
The extrinsic curvature terms in Eq.~(\ref{halfway}) can be 
written in terms 
of the gravitational momentum according to the relationship 
\be 
N\sqrt{h} \bigl[ K_{\mu\nu}K^{\mu\nu} - (K)^2 \bigr]/(2\kappa) = 
P^{ij} {\dot h}_{ij} - 2 P^{ij} D_i V_j - 
(\kappa N/\sqrt{h}) \Bigl[ 2 P^{ij} P_{ij} - (P)^2 \Bigr] 
\ , 
\ee 
which reduces to an identity when the definition 
Eq.~(\ref{momenta}a) for $P^{ij}$ and the kinematical expression
\be
K_{ij} = -\frac{1}{2N}({\dot h}_{ij} 
- 2 D_{{\sss (}i} V_{j{\sss )}} ) \label{Kishdot} 
\ee
are used.  From the 
results derived in Appendix A, Eq.~(\ref{otherflows}a) in particular, 
the term involving the 
gradient of $\theta$ can be written as 
\be 
\barN\baru\cdot\nabla\theta = (t^\mu - \sigma^\mu_\nu V^\nu) \nabla_\mu\theta 
= {\dot\theta} - V^a \partial_a\theta \ .\label{gradtheta} 
\ee 
Putting these results together and performing an integration by 
parts on the $\dot{\theta}$ term, we find 
\be 
S = \int_\M d^4x \Bigl\{ P^{ij} {\dot h}_{ij} - N\H - V^i \H_i \Bigr\} 
- \int_\barT d^3x
\Bigl\{ (\theta/\kappa) \dot{\sqrt{\sigma}\,\,\,\,\,}\!\!\!\!
+ \sqrt{\sigma}\Bigl( \barN \barepsi - \barV^a \barj_a\Bigr)\Bigr\} 
\,\label{canaction} 
\ee 
where $\H$ and $\H_i$ are the Hamiltonian and momentum constraints, 
respectively. In the $\barT$ term of $S$ above, $\barN = N/\gamma$, 
$\barV^a = V^a$, and $\barepsi$ and $\barj_a$ are  given by 
Eqs.~(\ref{bardensities}a,b). 

An alternative expression for the boundary terms of $S$ can be obtained as 
by using the kinematical equation (\ref{Kishdot}) to rewrite 
the quantity
$\dot{\sqrt{\sigma}\,\,\,\,\,}\!\!\!\!
 = \partial\sqrt{\sigma}/\partial
t$. Projecting Eq.~(\ref{Kishdot}) 
onto $B$, we have the result 
$ \ell_{\mu\nu} = -  \sigma_\mu^\alpha \sigma_\nu^\beta 
({\dot\sigma}_{\alpha\beta} - 2 D_\alpha V_\beta )/(2N) $, 
whose trace  yields 
$ \partial\sqrt{\sigma}/\partial t = \sqrt{\sigma} 
( - N \ell + \sigma^{ij} D_i V_j ) $. In the last term of this expression, 
$\sigma^{ij}D_i V_j$ can be simplified by 
splitting $V_j$ into its normal and tangential parts. This yields 
$\sigma^{ij} D_i (\sigma_{jk} V^k + Nv n_j) = d_a V^a - Nvk$, and 
results in the useful expression 
\be 
\partial\sqrt{\sigma}/\partial t = 
\sqrt{\sigma}(\kappa \barN \barjperp +  d_a V^a) \label{sigmadot} 
\ee
for the time derivative of $\sqrt{\sigma}$.
Putting these changes together, we find that the 
action (\ref{canaction}) equals 
\be 
S = \int_\M d^4x \Bigl\{ P^{ij} {\dot h}_{ij} - N\H - V^i \H_i \Bigr\} 
- \int_\barT d^3x \sqrt{\sigma} \Bigl\{ \barN \barepsi - V^a j_a 
+\barN\barjperp\theta \Bigr\} \ .\label{canaction2} 
\ee 
In both forms (\ref{canaction}) and (\ref{canaction2}) for the 
action, the independent variables are 
$h_{ij}$, $P^{ij}$, $N$, and $V^i$. 

The variation of the action (\ref{canaction}) or (\ref{canaction2}) 
can be computed explicitly, although the calculation is
difficult,${}^{12}$
and the result is
\bea 
\delta S & = & (\hbox{terms that give the canonical equations of motion}) 
\nonumber\\ 
& & + \int^{t''}_{t'} d^3x\, P^{ij} \delta h_{ij} - \frac{1}{\kappa} 
\int_{B'}^{B''} d^2x\theta\delta\sqrt{\sigma}  \nonumber\\
& & + \frac{1}{\kappa}\int_{\barT}d^3x\sqrt{\sigma}
\Bigl\{\kappa \barN \barjperp + d_{a} V^{a} - (1/2)\sigma^{ab}
\dot{\sigma}{}_{ab}\Bigr\}\delta\theta\nonumber\\ 
& & + \int_{\barT} d^3x \sqrt{\sigma} \Bigl\{ -\barepsi \delta\barN 
+ \barj_a\delta\barV^a + (\barN/2) \bars^{ab}\delta\sigma_{ab} \Bigr\} 
\ . \label{varycanaction}
\eea
This is not an unexpected expression, in view of Eq.~(\ref{varyTraceK})
and the definitions (\ref{bardensities}). Notice that $\theta$ need not
be held fixed in the canonical variation principle, as the term which
multiplies $\delta\theta$ in (\ref{varycanaction}) is (\ref{sigmadot})
which vanishes as a consequence of the canonical equations of motion.
We remark that in obtaining the result (\ref{varycanaction}) from 
(\ref{canaction}) or (\ref{canaction2}), one must 
use the kinematical relations (\ref{gradtheta}) and
(\ref{sigmadot}). These
relations are included 
among the equations of motion.

\subsection{Variation of the Hamiltonian without boundary terms} 
The ``base" Hamiltonian for general relativity, unaugmented by boundary
terms, is 
\be 
H_{\rm base} = \int_\Sigma d^3x \bigl( N\H + V^i \H_i \bigr) \ ,
\label{basehamiltonian}
\ee 
where  the Hamiltonian and momentum constraints are 
\bea 
\H &=& \frac{\kappa}{\sqrt{h}}(2P^{ij}P_{ij} - P^2) - 
\frac{1}{2\kappa}\sqrt{h} R 
\ ,\\
\H_i &=& -2D_j P^j_i \ .
\eea
For convenience, we write $H_{\rm base} = H_{N} + H_{\vec{V}}$, where e.~g.~we
define $H_{N} \equiv \int_{\Sigma}d^3x N\H$.
In the calculation of $\delta H$ below, we only keep terms that give 
rise to boundary terms. This avoids clutter in our presentation, and
in any case these are the difficult terms to isolate correctly in 
the variation.

First consider the smeared Hamiltonian constraint, denoted $H_N$. 
We have 
\be 
\delta H_N = \cdots -\frac{1}{2\kappa} \int_\Sigma d^3x\, N\sqrt{h}
h^{ij}\delta R_{ij} 
\ ,\label{smearedHN}
\ee 
where the dots denote terms that do not give rise to boundary terms. 
This calculation is nearly identical to the calculation of the variation of 
the Hilbert action from Section II. Thus, we find  
\bea 
h^{ij}\delta R_{ij} &=& D_i \Lambda^i \ ,\nonumber\\
\Lambda^i &=& 2h^{j{\sss [}i} D^{k{\sss ]}} \delta h_{jk} \ ,\nonumber 
\eea 
and Eq.~(\ref{smearedHN}) becomes 
\be
\delta H_N = \cdots +\frac{1}{2\kappa}\int_\Sigma d^3x \sqrt{h} (D_i
N)\Lambda^i 
- \frac{1}{2\kappa}\int_{\partial\Sigma} d^2x\sqrt{\sigma} N n_i\Lambda^i \ .
\label{smearedHN2} 
\ee 
Moreover, our proof of the lemma (\ref{lemma}) in Section II goes 
through unaltered for the case at hand (a lower
dimensional setting). Therefore, we have     
\be 
n_i \Lambda^i = 2\delta k 
+ k^{ij} \delta h_{ij} + d_i ( \sigma^i_j\delta n^j) \ , 
\ee 
where $d_i$ is the covariant derivative on $\partial\Sigma$. Now, 
the first term in Eq.~(\ref{smearedHN2}) involves 
\be
(D_i N)\Lambda^i = 
2 h^{j{\sss [}i} D^{k{\sss ]}} \bigl[ (D_i N) \delta h_{jk} 
\bigr] - 2h^{j{\sss [}i} \bigl[ 
D^{k{\sss ]}}(D_i N)\bigr] \delta h_{jk} \ , 
\ee 
so that, keeping only boundary terms, we find 
\bea
\delta H_N = \cdots + 
\frac{1}{2\kappa}\int_{\partial\Sigma} d^2x\sqrt{\sigma} 
\biggl\{ &-& 2N\delta k - Nk^{ij}
\delta h_{ij} + (d_i N) \delta n^i \nonumber\\
&+& n^i (D^j N) \delta h_{ij} - 
(n^i D_i N) h^{jk} \delta h_{jk} \biggr\} \ . 
\eea
With the substitution $h_{ij} = \sigma_{ij} + n_i n_j$ and 
the useful identities $n^i\delta\sigma_{ij} = - \sigma_{ij}\delta n^i$, 
$\delta n_i = n_i n^j \delta n_j$, and $h^{ij}\delta \sigma_{ij} 
= \sigma^{ij}\delta\sigma_{ij}$, one obtains 
\be 
\delta H_N = 
\cdots -\frac{1}{2\kappa} \int_{\partial\Sigma}
d^2x\sqrt{\sigma} 
\biggl\{ 2N\delta k + \bigl[ N k^{ij} 
+ n^k(D_k N)\sigma^{ij}\bigr] \delta\sigma_{ij} 
\biggr\} \label{finalvaryHN} 
\ee 
for the contribution to $\delta H$ from the $H_N$ term. 

Now consider the smeared momentum constraint, denoted 
$H_{\vec{V}}$. It is straightforward to show that 
\be 
\delta \H_i = 
-2D_j[\delta(P^{jk}h_{ki})] + P^{jk} D_i(\delta h_{jk}) \ , 
\ee 
from which one obtains
\be 
\delta H_{\vec{V}} = \cdots +\int_{\partial\Sigma} d^2x
(\sqrt{\sigma}/\sqrt{h}) \biggl\{ 
- 2n_i V_k \delta P^{ik}
+\bigl[ (n\cdot V) P^{jk} 
- 2n_i P^{ij} V^k \bigr]\delta h_{jk} \biggr\} \ .
\ee 
The first term in $\delta H_{\vec{V}}$ 
can be rewritten by noting that the factor 
$\sqrt{\sigma} n_i /\sqrt{h}$ is metric 
independent, and can be passed inside the variation $\delta$. With
the shorthand notation 
\be 
\jmath^k = -2n_i P^{ik}/\sqrt{h} \ ,\label{jmath} 
\ee 
the first term in the integrand of $\delta H_{\vec{V}}$ becomes 
$V_k\delta(\sqrt{\sigma} \jmath^k)$. 
The remaining terms in $\delta H_{\vec{V}}$ can be rewritten with the
result  
\bea 
\delta h_{jk} &=& h_j^\ell h_k^m \delta h_{\ell m} \nonumber\\
&=& 2n_j n_k n^\ell \delta n_\ell + \sigma_j^\ell \sigma_k^m 
\delta \sigma_{\ell m} + 2
\sigma^\ell_{{\sss (}j}  n_{k{\sss )}} 
n^m \delta\sigma_{\ell m}  \ ,
\eea 
which is derived by using the substitution 
$h_{ij} = \sigma_{ij} + n_i n_j$ and 
the useful identities mentioned previously. With these changes, 
we obtain 
\bea 
\delta H_{\vec{V}} = 
\cdots +\int_{\partial\Sigma} d^2x\biggl\{ &&
V_k\delta(\sqrt{\sigma} \jmath^k)
+ \sqrt{\sigma} (n\cdot V) \jmath^k n_k n^i \delta n_i \nonumber\\
&& +\sqrt{\sigma} \Bigl[
(n\cdot V) P^{k\ell} \sigma^i_k \sigma^j_\ell/\sqrt{h}  
+ (V^k\sigma_k^i) \jmath^\ell\sigma_\ell^j 
+ (V^k\sigma_k^i)  \jmath^\ell n_\ell n^j\Bigr]\delta\sigma_{ij} 
\biggr\} \ .
\eea
Our next task is to simplify the term 
$V_k\delta (\sqrt{\sigma}\jmath^k)$. Using the useful
identities,  we find 
\bea 
V_k \delta (\sqrt{\sigma}\jmath^k )
&=& V_i(\sigma^i_k + n^i n_k) \delta (\sqrt{\sigma} \jmath^\ell(\sigma_\ell^k + n_\ell n^k))
\nonumber\\  &=& (V^k \sigma^i_k) \delta(\sqrt{\sigma}\jmath^\ell\sigma_{\ell i}) 
+ (n\cdot V) \delta (\sqrt{\sigma}\jmath^k n_k)
 -  \sqrt{\sigma}(n\cdot V) \jmath^kn_k n^i\delta n_i  \nonumber\\ 
&&  - \sqrt{\sigma}(V^k\sigma^i_k)  \jmath^\ell\sigma_\ell^j\delta\sigma_{ij} - 
 \sqrt{\sigma} (V^k\sigma_k^i)\jmath^\ell n_\ell n^j \delta \sigma_{ij} 
\ .
\eea 
The last three terms in $V_k \delta (\sqrt{\sigma}\jmath^k )$ cancel other 
terms in the integrand of $\delta H_{\vec{V}}$, leaving us with 
\bea
\delta H_{\vec{V}} = \cdots \int_{\partial\Sigma} d^2x \biggl\{ (n\cdot
V) (\sqrt{\sigma} 
\sigma^i_k \sigma^j_\ell P^{k\ell} /\sqrt{h}) \delta\sigma_{ij} 
&-& 2 (V^\ell \sigma_\ell^i)\delta(\sqrt{\sigma} n_j P^{jk}\sigma_{ki}/\sqrt{h}) 
\nonumber\\
&-& 2(n\cdot V) \delta( \sqrt{\sigma} n_i P^{ij} n_j /\sqrt{h}) \biggr\} 
\ .\label{finalvaryHV}
\eea
Here, the definition (\ref{jmath}) has been used to express $\jmath^k$ in
terms of $P^{ij}$. 

Collecting the results from Eqs.~(\ref{finalvaryHN}) and 
(\ref{finalvaryHV}), we have 
\bea 
\lefteqn{\delta H_{\rm base}} & & \nonumber
\\ 
& = & \cdots
+\int_{\partial\Sigma} d^2x \biggr\{
-\frac{1}{\kappa} N\sqrt{\sigma} \delta k 
- 2(n\cdot V) \delta(\sqrt{\sigma} 
n_i P^{ij} n_j /\sqrt{h}) 
-2V^\ell\sigma_\ell^i \delta(\sqrt{\sigma} 
n_j P^{jk}\sigma_{k\ell}/\sqrt{h} ) \biggr.
\nonumber\\ 
& & \biggl. -\sqrt{\sigma} \biggl[ \frac{1}{2\kappa} 
[Nk^{ij} + n^k(D_k
N)\sigma^{ij}] 
  -(n\cdot V) \sigma^i_k \sigma^j_\ell 
P^{k\ell}/\sqrt{h} \biggr] \delta\sigma_{ij} 
\biggr\} \ .
\eea 
We can rewrite this expression in terms of coordinates $x^a$ on the surface 
$\partial\Sigma$: Let $\partial\Sigma$ 
correspond to an $r={\rm const}$ surface, and define  
$\sigma_a^i = \partial x^i(x^a,r)/x^a$. Then  $\delta\sigma_a^i = 0$. Also let 
$\sigma^{ij} = \sigma^{ab}\sigma_a^i \sigma_b^j$, 
$V^i = V^a \sigma^i_a + (n\cdot V) n^i$, and $k^{ij} = k^{ab} \sigma_a^i \sigma_b^j$. 
Then we have 
\bea 
\delta H_{\rm base} = \cdots +\int_{\partial\Sigma} d^2x \biggr\{ &&
-\frac{1}{\kappa} N\sqrt{\sigma} \delta k 
- 2(n\cdot V) \delta(\sqrt{\sigma} n_i P^{ij} n_j /\sqrt{h}) 
-2V^a \delta(\sqrt{\sigma} n_j P^{jk}\sigma_{ka}/\sqrt{h} ) 
\nonumber\\ 
&& -\sqrt{\sigma} \biggl[ \frac{1}{2\kappa} [Nk^{ab} + n^k(D_k N)\sigma^{ab}] 
  -(n\cdot V) \sigma^a_k \sigma^b_\ell P^{k\ell}/\sqrt{h} \biggr] \delta\sigma_{ab} 
\biggr\} \ .
\eea 
In terms of the boost velocity $v = (n\cdot V)/N$ and the quasilocal densities 
\bea 
\epsi &=& k/\kappa \ ,
\label{dens}\eqnum{\ref{dens}a}\\
\jperp &=& -2n_i P^{ij} n_j/\sqrt{h} \ ,\eqnum{\ref{dens}b}\\
\jmath_a &=& -2 \sigma_{ai} P^{ij} n_j/\sqrt{h} \ ,\eqnum{\ref{dens}c}\\ 
s^{ab} &=& \frac{1}{\kappa} \biggl( k^{ab} + \biggl[ \frac{n^i \partial_i N}{N} 
- k \biggr]\sigma^{ab} \biggr) \ ,\eqnum{\ref{dens}d}\\
t^{ab} &=& 2\sigma^a_i \sigma^b_j P^{ij}/\sqrt{h} \ ,\eqnum{\ref{dens}e} 
\addtocounter{equation}{1}
\eea 
we obtain 
\be 
\delta H_{\rm base} = \cdots -\int_{\partial\Sigma} d^2x \biggl\{ 
N\Bigl[ \delta(\sqrt{\sigma}\epsi) - v\delta(\sqrt{\sigma}\jperp)\Bigr] 
- V^a\delta(\sqrt{\sigma} \jmath_a)  
+ (N\sqrt{\sigma}/2) (s^{ab} - vt^{ab})\delta\sigma_{ab} \biggr\} 
\ee 
for the boundary terms in the variation of the base gravitational Hamiltonian.

\subsection{Boost Relations for $\epsi$, $\jperp$, and $\jmath_a$ from
the Hamiltonian}

The gravitational Hamiltonian $H[N,V]$ whose values are the 
quasilocal energy density $\epsi$ and quasilocal momentum density 
$\jmath_i$ is \be 
H[N,V] = \int_\Sigma d^3x \Bigl( N\H + V^i \H_i \Bigr) 
+ \int_{\partial\Sigma} d^2x \sqrt{\sigma} 
\Bigl( N\epsi - V^i \jmath_i\Bigr) \ .
\label{TheHamiltonian}
\ee
The variation of this Hamiltonian with respect to the 
canonical variables is 
\be 
\delta H[N,V] = \int_\Sigma d^3x \biggl\{ \frac{\delta H[N,V]}{\delta h_{ij}} 
\delta h_{ij} + \frac{ \delta H[N,V]}{\delta P^{ij}} \delta P^{ij} \biggr\} 
+ \delta H[N,V]\Bigr|_{\partial\Sigma} \ ,
\ee 
where 
\bea 
\frac{\delta H[N,V]}{\delta h_{ij}} &=& -\frac{\kappa N}{\sqrt{h}} (P^{k\ell}P_{k\ell} 
- P^2/2) h^{ij} + \frac{2\kappa N}{\sqrt{h}} (2P^{ik}P_k^j - PP^{ij}) \nonumber\\
&& +\, \frac{\sqrt{h}N}{2\kappa} G^{ij} - \frac{\sqrt{h}}{2\kappa} D^iD^jN + 
\frac{\sqrt{h}}{2\kappa} (D_k D^k N) h^{ij} \nonumber\\ 
&&\, + 2 (D_kV^{{\sss (}i} ) P^{j{\sss )}k} - D_k(P^{ij} V^k)  \ ,
\label{varyHNV}\eqnum{\ref{varyHNV}a}\\
\frac{ \delta H[N,V]}{\delta P^{ij}} &=& \frac{2\kappa N}{\sqrt{h}} (2P_{ij} - Ph_{ij}) 
+ 2D_{{\sss (}i} V_{j{\sss )}} \ ,\eqnum{\ref{varyHNV}b} 
\addtocounter{equation}{1}
\eea 
and the boundary terms are
\be
\delta H[N,V]\Bigr|_{\partial\Sigma} = \int_{\partial\Sigma} d^2x \sqrt{\sigma} \Bigl[
\epsi \delta N -  \jperp \delta(Nv) - \jmath_a \delta V^a 
- (N/2) (s^{ab} - vt^{ab})\delta\sigma_{ab} \Bigr] \ .\label{varyHbts} 
\ee
The terms are readily found using the results obtained in the last subsection for
the variation of $H_{\rm base}$. Now let us compute the change in the 
Hamiltonian corresponding
to a quasilocal boost. That is, perform a surface deformation that 
becomes an infinitesimal pure boost at the boundary
$\partial\Sigma$ (or, more precisely, becomes an infinitesimal pure boost in the
orthogonal complement to the tangent space of each boundary point). The surface
deformation is 
described by a deformation vector, which 
we  split into a normal part (lapse function) $\eta$ and a tangential part
(shift vector) $\nu^i$. The characteristics of an infinitesimal boost at 
$\partial\Sigma$ are 
\bea 
\eta\bigr|_{\partial\Sigma} &=& 0 \ ,
\label{surfacedef}\eqnum{\ref{surfacedef}a}\\
\nu^i\bigr|_{\partial\Sigma} &=& 0 \ ,\eqnum{\ref{surfacedef}b}
\addtocounter{equation}{1}
\eea 
and 
\be 
n^i\partial_i \eta\bigr|_{\partial\Sigma} = \dot\theta  \ ,\label{roctheta}
\ee 
where $\theta$ is the velocity parameter (see Appendix C for an explanation of 
the relevant geometry of this assignment). Under a
surface 
deformation, the changes in the canonical variables are 
\bea 
\delta h_{ij} &\equiv& {\dot h}_{ij} dt 
= \frac{\delta H[\eta,\nu]}{\delta P^{ij}} dt \ ,\\
\delta P^{ij} &\equiv& {\dot P}^{ij} dt 
= - \frac{\delta H[\eta,\nu]}{\delta h_{ij}} dt \ ,
\eea 
where $\delta H/\delta P^{ij}$ and $\delta H/\delta h_{ij}$ are given by 
Eq.~(\ref{varyHNV}). 

A surface deformation only affects the canonical variables. 
By definition, the lapse and shift remain unchanged, $\delta N = 0 = \delta V^i$. 
In the surface terms of Eq.~(\ref{varyHbts}), we must consider the variations of $Nv$, $
V^a$, 
and $\sigma_{ab}$. 
First, let us look at $\delta(Nv)$: 
\bea 
\delta(Nv)\bigr|_{\partial\Sigma} &=& \delta(n_i V^i)\bigr|_{\partial\Sigma} \nonumber\\
&=& (n_i n^j n^k \delta h_{jk}/2) V^i\bigr|_{\partial\Sigma} \nonumber\\
&=& \frac{1}{2} (n\cdot V) n^j n^k \frac{\delta H[\eta,\nu]}{\delta P^{jk}} 
dt\biggr|_{\partial\Sigma} \nonumber\\ 
&=& Nv n^i n^j (D_j \nu_i) dt\bigr|_{\partial\Sigma} \ . 
\eea 
In deriving this result, we have used Eqs.~(\ref{varyHNV}b) 
and (\ref{surfacedef}a). Now turn to
the variation 
of $V^a$. Recall that $V^a = V^i \sigma_i^a$ with  $\sigma_i^a = \sigma^{ab} h_{ij} \sigma_b^j$, 
$\sigma_{ab} = \sigma_a^i h_{ij} \sigma_b^j$, and  with $\sigma_b^j$ metric independent. 
A calculation similar to the one above, which uses the conditions (\ref{surfacedef}), yields 
\be 
\delta V^a \bigr|_{\partial\Sigma} 
= Nv \sigma^{ai} n^j (D_j \nu_i) dt \bigr|_{\partial\Sigma} \ . 
\ee 
Although it would seem to us not necessary, we find it convenient to choose 
the shift part of the deformation, $\nu^i$, so that $\delta (Nv) = 0 = \delta V^a$. 
Therefore, we impose 
\be 
n^j D_j \nu_i\bigr|_{\partial\Sigma} = 0 \ ,\label{gaugechoice} 
\ee 
as an additional condition, along with Eq.~(\ref{surfacedef}). Finally, consider 
the variation of $\sigma_{ab}$. It is not difficult  to show that 
\be
\delta\sigma_{ab}\bigr|_{\partial\Sigma} 
= 2 d_{{\sss (}a} \nu_{b{\sss )}} \bigr|_{\partial\Sigma} \ , 
\ee 
where $d_a$ is the covariant derivative on $\partial\Sigma$. Since 
$\nu^i$ vanishes on $\partial\Sigma$, we find that $\delta \sigma_{ab} = 0$ 
on $\partial\Sigma$. 
This, along with the results $\delta N = 0$, $\delta(Nv) = 0$, and 
$\delta V^a = 0$ on $\partial\Sigma$, implies that the boundary term 
(\ref{varyHbts}) is zero under the variation defined by the boost $\eta$, $\nu^i$. 

The results above show that the variation in the Hamiltonian  is 
$\delta H[N,V] \equiv {\dot H} dt$ where 
\be
\dot H = \int_\Sigma d^3x \biggl\{ \frac{\delta H[N,V]}{\delta h_{ij}} 
\frac{\delta H[\eta,\nu]}{\delta P^{ij}} - \frac{\delta H[N,V]}{\delta P^{ij}} 
\frac{\delta H[\eta,\nu]}{\delta h_{ij}} \biggr\} \ .\label{rocH}
\ee 
We comment on this equation in more detail below.
The next step is to insert the results from Eq.~(\ref{varyHNV}) into 
the expression (\ref{rocH}) for $\dot H$ and simplify. 
Although the calculation is essentially straightforward, it is also somewhat 
long and difficult. The result is 
\bea 
{\dot H}[N,V] &=& \int_\Sigma d^3x \biggl\{ {\dot N}\H 
+ {\dot V}^i\H_i \biggr\} \nonumber\\
& & + \int_{\partial\Sigma} d^2x \sqrt{\sigma} \biggr\{ 
\frac{2}{\sqrt{h}} (ND_i\eta)P^{ij}n_j + \frac{1}{\kappa}(n^iD_i N) d_a\nu^a 
- \frac{1}{\kappa}(n^iD_i N)(n\cdot\nu)k \nonumber\\
& & \qquad\qquad\qquad - \frac{1}{\kappa} (D_i N) k^{ij}\nu_j 
- \frac{1}{\kappa}(D^i N)d_i(n\cdot\nu) 
- \frac{1}{\sqrt{h}} \eta(n\cdot V) \H \nonumber\\
& & \qquad\qquad\qquad + \frac{1}{\kappa} N n_i R^{ij}\nu_j 
+ \frac{2}{\sqrt{h}} V^i P^{jk} \Bigl[ n_{{\sss [}k} d_{i{\sss ]}} 
(\sigma_j^\ell \nu_\ell) - (n\cdot\nu) n_{{\sss [}k} k_{i{\sss ]}j} 
\nonumber\\ 
& & \qquad\qquad\qquad + n_{{\sss [}k} k_{i{\sss ]}\ell} \nu^\ell n_j 
+ n_{{\sss [}k} d_{k{\sss ]}}(n\cdot\nu) n_j \Bigr] 
- (N,V \leftrightarrow \eta,\nu)\biggr\} \ , 
\eea 
where we have defined 
\bea 
{\dot N} &\equiv& V^i D_i\eta - \nu^i D_i N \ ,\\
{\dot V}^i &\equiv& ND^i \eta + V^jD_j\nu^i - \eta D^i N - \nu^j D_j V^i \ . 
\eea 
Next, we impose the boundary conditions (\ref{surfacedef}) [note, we do not need to use
Eq.~(\ref{gaugechoice})], and obtain
\bea 
{\dot H}[N,V] &=& \int_\Sigma d^3x \biggl\{ {\dot N}\H 
+ {\dot V}^i\H_i \biggr\} \nonumber\\
& & + \int_{\partial\Sigma} d^2x \sqrt{\sigma} (n^i D_i\eta) \Bigl[ 
2 N n_i P^{ij}n_j/\sqrt{h}
+ (n\cdot V)k/\kappa  -   d_aV^a /\kappa \Bigr] \ . 
\eea 
With the definitions (\ref{dens}) of the quasilocal energy and momentum densities,
one then writes 
\be 
{\dot H}[N,V] = \int_\Sigma d^3x \biggl\{ {\dot N}\H 
+ {\dot V}^i\H_i \biggr\} 
+ \int_{\partial\Sigma} d^2x \sqrt{\sigma} \Bigl[ 
-N{\dot\theta} \jperp + Nv{\dot\theta}\epsi 
+ V^a\partial_a{\dot\theta}/\kappa \Bigr] \ , \label{rocH2}
\ee 
where $\theta$ is the velocity parameter defined in Eq.~(\ref{roctheta}). 

When the constraints hold, $\H = 0 = \H_i$, the energy surface density 
$\epsi$ and momentum surface density $\jmath_i$  are the values of the 
Hamiltonian $H[N,V]$ associated with various lapse functions $N$ and shift 
vectors $V^i$. That is, the coefficients of $N$ and $V^i$ in the surface 
terms of $H$ are $\epsi$ and $\jmath_i$, respectively. The expression for 
$\dot H$ above gives the change in $H$ under a surface deformation that 
becomes a boost at the boundary $\partial\Sigma$. From Eq.~(\ref{rocH2}) 
we determine the changes in the values of $H$ under a boost, namely
\bea 
\dot\epsi &=& - {\dot\theta}\jperp \ ,
\label{rocdens}\eqnum{\ref{rocdens}a}\\
\dot\jperp &=& -{\dot\theta}\epsi \ ,\eqnum{\ref{rocdens}b}\\
\dot\jmath_a &=& -\partial_a{\dot\theta}/\kappa \ .\eqnum{\ref{rocdens}c} 
\addtocounter{equation}{1}
\eea 
These expressions can be integrated to obtain the boost relations for a 
finite boost, as follows. The first two equations, 
\bea 
\frac{d\epsi}{d\theta} &=& - \jperp \ ,\\
\frac{d\jperp}{d\theta} &=& - \epsi \ , 
\eea
for the energy and normal momentum surface densities, have the solution 
\bea 
\epsi(\theta) &=& \gamma\, \epsi(0) - \gamma v\, \jperp(0) \ ,\\ 
\jperp(\theta) &=& \gamma\, \jperp(0) - \gamma v\, \epsi(0) \ ,
\eea 
where $\gamma = \cosh\theta$ and $\gamma v = \sinh\theta$. Similarly, 
Eq.~(\ref{rocdens}b) 
yields 
\be 
\jmath_a(\theta) = \jmath_a(0) - \partial_a\theta/\kappa  
\ee 
for the tangential components of the momentum surface density. These results 
are equivalent to the boost relations (\ref{epjboosts}) and (\ref{tanjboost}).


\section{Examples of Quasilocal Energy and Momentum}
We now examine our quasilocal energy and momentum for several spherically 
symmetric scenarios, including static solutions of Einstein's equations, 
a boosted foliation of the Schwarzschild geometry, and isotropic 
cosmologies. (Dadhich and Bose have used the quasilocal energy and
the gravitational charge defined by the Komar integral to
characterize black--hole horizons for spherically
symmetric spacetimes.\cite{Dadhich})
We then examine a non-spherically symmetric scenario, namely
cylindrical gravitational waves. These examples, and our treatment of
energy--momentum at spatial infinity in Sec.~VI, require that we 
address the issue of zero--points for quasilocal energy--momentum, and 
we begin with a discussion of this issue. We do not present concrete examples
of quasilocal stress, however we draw attention to Ref.~\cite{BC} in which
quasilocal stress was considered by Booth and Creighton in their calculation 
of the tidal heating of Jupiter's moon Io.

\subsection{Subtraction term}
For any variational principle one has the freedom to add to the action terms 
that depend on the fixed boundary data. 
Thus, we can append  a ``subtraction term" $-S^0$ 
to the Trace-K action $S$, which is a functional $-S^0[\bargam_{ij},h_{ij}]$ 
of the fixed boundary data 
$\bargam_{ij}$ and $h_{ij}$. The modified action $S - S^0$, like $S$ 
itself, yields the Einstein equations as equations of motion when varied 
subject to fixation of $\bargam_{ij}$ and $h_{ij}$ on the spacetime 
boundary $\partial\M$. Certain modifications are brought about
in redefining the action $S \to S - S^0$ to include a subtraction term; 
however, before turning to the details of these modifications, let us 
point out that ---as indicated in the fourth footnote of the introduction--- 
the freedom associated with the subtraction term implies that our formalism
alone does not completely determine a definition for quasilocal energy
and related quantities.
This ambiguity is a field--theoretic version of the standard one
associated with any finite dimensional mechanical system described by a 
variational principle, namely the freedom both to choose the 
zero--point value of the energy and to redefine the system's momenta 
via canonical transformation. (Ref.~\cite{BY} spells out the analogy
between the field--theoretic freedom present in the {\sc cqf} and the
corresponding freedom in finite--dimensional mechanics in some 
detail.${}^{13}$)
As such, the subtraction term is not a background structure {\em per se}.
However, we may and often do in practice introduce a background structure,
a ``reference space," as a vehicle for introducing a particular
physically relevant subtraction term, and we therefore use the terms
``reference'' and ``zero--point'' as synonyms.

The freedom to include a subtraction term in the action leads to modified
expressions for quasilocal energy--momentum, and ones which are defined
uniquely only up to reference contributions. In the interest
of economy, let us confine our comments (for the moment) to the modified 
expression $\kappa\bar{\varepsilon} = \bar{k} - \bar{k}^0$ 
for the quasilocal energy surface density,
where $\bar{k}^0$ stems from the inclusion
of $-S^0$. It should be clear that
our discussion also pertains to the momentum densities $\bar{\jmath}_k$. 
The issue then is how to resolve the ambiguity in the formalism by selecting a
suitably unique reference term $\bar{k}^0$. As we show later in Sec.~VI, if the 
goal is to obtain agreement between the {\sc cqf} and the accepted notions 
of total gravitational energy (either at spatial or null infinity), then 
there is a suitably unique choice of energy zero-point\cite{BY,BLY1,BLY2,Lau4}. 
However, at the quasilocal level there is no known preferred choice, other 
than the choice $\bar{k}^0 = 0$. (This simple choice has proven 
useful in itself, as seen in Sec.~III.C and more recently in an approach to 
numerical outer boundary conditions.\cite{ArrestedHistory} However, it 
leads to an infinite energy in large sphere limits.) At first sight it 
would seem that the the zero--point ambiguity in the {\sc cqf} is no better 
or worse than the situation encountered with, say, pseudo--tensor descriptions 
of gravitational {\sc sem}, which are plaqued by ambiguity in the choice of 
background coordinates (or more generally in the choice of moving frame). 
However, the {\sc cqf} offers some new insight into the ambiguous nature of 
gravitational {\sc sem}. Foremost, it provides a physical interpretation 
for the ambiguity, and one that is a field-theoretic generalization of the
standard ambiguity present in the Hamilton--Jacobi description of ordinary 
mechanics. Moreover, as we discuss further below, in our formalism the 
selection of zero--point is usually cast as an embedding problem, affording a
precise mathematical interpretation in terms of an associated {\sc pde} problem.

These insights aside, let us state again that we do not have an over-arching 
rule, applicable for all quasilocal two--surfaces, for selecting (suitably 
unique) zero-points. In our view it is the physicist's job to select the 
appropriate choice of zero-point on a case-by-case basis, with the only guide 
being the rather nebulous principle that the selection should be tailored to 
the ``physics'' of the scenario at hand. We would like to point out that this 
is a common enough state of affairs in general relativity, a {\em meta} theory 
known for its wealth of possible boundary conditions. Indeed, by way of analogy 
consider the search for solutions of the Einstein field equations. In practice, 
relativists certainly do not attempt to find {\em the} general solution, rather 
they attempt to find solutions given some additional physical input (boundary 
conditions, symmetries, etc.). In practice the same such additional input is 
needed to associate a meaningful {\sc qle} with a particular quasilocal 
two--surface. These considerations suggest that, rather than albatross, 
the zero--point ambiguity is a desirable feature of the {\sc cqf}, as it 
affords pliable enough definitions of {\sc sem} to have broad application in 
general relativity.

Turning now to the technical details,
let us note that Eqs.~(\ref{bardensities}) and (\ref{newbar}), along with 
the unbarred versions of Eqs.~(\ref{newbar}), yield the purely kinematical 
relationships 
\bea 
\bark &=& \gamma k + \gamma v \ell \ ,
\label{A.1}\eqnum{\ref{A.1}a}\\
\barell &=& \gamma \ell + \gamma v k \ ,\eqnum{\ref{A.1}b}\\
\sigma_a^i\barK_{ij}\barn^j &=& \sigma_a^i K_{ij} n^j -
\partial_a\theta \ ,\eqnum{\ref{A.1}c}
\addtocounter{equation}{1}
\eea 
with similar relations stemming from Eqs.~(\ref{bardensities}c,f). 
The inclusion of a subtraction term, $S \to S - S^0$, will modify the 
definition (\ref{bardensities}a) so that 
$\kappa\barepsi = \bark -\bark^0$.  Here, as suggested in the original 
paper\cite{BY}, we have chosen the subtraction term $S^0$ such that the 
quasilocal energy surface density acquires a term $\bark^0$ which
is the trace of the extrinsic curvature of a surface $B$ with metric 
$\sigma_{ab}$ embedded in some reference space. 
(Note that given a reference space, a spacelike slice of some
fixed spacetime with boundary metric equal to $\sigma_{ab}$, a
family of reference spaces can be generated by boosting the slice
at the surface $B$.)
This requires the $\barT$ contribution to 
$S^0$ to be  a linear functional of $\barN$ with coefficient 
$-\sqrt{\sigma}\bark^0/\kappa$. Likewise, by choosing the 
$\Sigma''$ contribution to $S^0$
to be a linear functional of $\barM$ with an 
appropriate coefficient, we obtain a
modified version of Eq.~(\ref{bardensities}d),  
namely $\kappa\barjperp =
-\barell + \barell^0$. Finally, 
by choosing the $\barT$ contribution to $S^0$ to be a linear functional 
of $\barV^a$ with an appropriate coefficient (and choosing the $\Sigma''$ 
contribution to $S^0$ to be a linear functional of $\barW^a$ with an
appropriate 
coefficient), we obtain a 
modified version of  Eq.~(\ref{bardensities}b,e), namely
$\kappa\barj_a =
\sigma_a^i \barK_{ij} \barn^j 
- (\sigma_a^i \barK_{ij} \barn^j)|^0$. How do these modifications 
of $\barepsi$, $\barjperp$, and $\barj_a$ affect the boost
relations (\ref{epjboosts}), (\ref{tanjboost})? The answer 
depends on the relationship between the subtraction terms for
different observers. 
If the subtraction terms $k^0$, $\ell^0$, and $(\sigma^i_a K_{ij} n^j)|^0$ for 
different observers are chosen such that they  
satisfy the kinematical relationships (\ref{A.1}), then the boost 
relations become 
\bea 
\barepsi &=& \gamma \epsi - \gamma v \jperp \ ,
\label{A.2} \eqnum{\ref{A.2}a}\\
\barjperp &=& \gamma \jperp - \gamma v \epsi \ ,
\eqnum{\ref{A.2}b}\\
\barj_a &=& \jmath_a \ , \eqnum{\ref{A.2}c} 
\addtocounter{equation}{1}
\eea 
where $\kappa\epsi = k - k^0$, $\kappa\jperp = -\ell + \ell^0$, and 
$\kappa\jmath_a = \sigma_a^i  K_{ij}  n^j - (\sigma_a^i  K_{ij}  n^j)|^0$. 
In order to have $k^0$, $\ell^0$, and $(\sigma_a^i \barK_{ij} \barn^j)|^0$ 
related as in Eqs.~(\ref{A.1}), we must choose a fiducial reference space
for the subtraction term for some fixed observers (say, the unbarred observers), 
then choose the reference space for the subtraction term for all other 
observers to be boosted relative to the fiducial reference space. Here we do 
not discuss modifications of $s^{ab}$ and $t^{ab}$ arising from a subtraction 
term, but see Ref.~\cite{Lau3} where these stresses are modified in an 
Ashtekar--variable reformulation of portions of this theory.

The construction of reference term $k^0$ from a reference space amounts
to posing and solving an isometric embedding problem. One natural choice
---discussed in the original paper Ref.~\cite{BY}--- is to embed $B$ isometrically
into Euclidean three-space $\referenceE^3$ in order to obtain an
extrinsic curvature tensor $(k^0)_{ab}$ (and hence $k^{0}$), a task tantamount 
to solving the following system of {\sc pde}:
\bea
(k^{0}){}^{2} - (k^{0})_{ab} (k^{0}){}^{ab} - {\cal R} & = & 0 
\label{Weylproblem} \eqnum{\ref{Weylproblem}a}\\
d_{b} k^{0} - d_{a} (k^{0}){}^{a}{}_{b} & = & 0\, ,
\eqnum{\ref{Weylproblem}b}
\addtocounter{equation}{1}
\eea
where $\cal R$ is the $B$ Ricci scalar and the $d_{a}$ denotes covariant
differentiation in the $B$ metric. 
These are the Gauss-Codazzi-Mainardi constraint equations for
$B$ embedded in $\referenceE^{3}$, and may be derived as in
Ref.~\cite{York}.   
Notice that $(k^0)_{ab}$ is indeed determined
solely by $\sigma_{ab}$ (albeit non-locally in general). With the second fundamental
form $(k^{0})_{ab}$ expressed in terms of the embedding's coordinate chart, this 
is {\em Weyl's problem},
a classic problem of differential geometry in the large for which an extensive
literature exists. In a somewhat recent formulation of the problem, Heinz\cite{Heinz}
has proven the existence of such an embedding if the $B$ scalar
curvature $\cal R$ is everywhere positive and the metric functions $\sigma_{ab}$ are of 
$C^2$ differentiability class. Uniqueness of the embedding, up to Euclidean motions,
then follows from the ``rigidity theorem'' of Cohn-Vosson.\cite{Spivak}
While such a Euclidean or ``flat-space'' reference proves important when one
considers asymptotic limits of the quasilocal energy, we note that other useful
prescriptions for referencing quasilocal energy can be devised; see for example
Refs.~\cite{BLY2,Lau4,Epp}. In these works, the idea has been to use a non--inertial
hyperplane of Minkowski spacetime as the reference space. Note that such non--inertial
slices do not have non-vanishing extrinsic curvature tensors, and hence would give
rise to reference contributions to the momentum densities $\jmath_k$. That said, in those
works the main focus has been on the energy expression.

In what follows, we shall assume that the quasilocal energy surface density 
$\varepsilon = (k - k^{0})/\kappa$ has been set via Euclidean reference. 
This means that for a fixed metric geometry $\sigma_{ab}$ on $B$, the total 
quasilocal energy
\be
E = \frac{1}{8\pi}
\int_{B} d^{2}x \sqrt{\sigma} \left(k - k^{0}\right)
\label{totalQLE}
\ee
arises as the difference between two total mean curvatures (here we set 
Newton's constant to unity so that $\kappa = 8\pi$). The first,
the proper integral of $k/(8\pi)$, is associated with the embedding of
$B$ in a hypersurface $\Sigma$ of the physical spacetime. The second, the 
proper integral of $k^{0}/(8\pi)$, is associated with an {\em isometric} 
embedding of $B$ in an auxiliary Euclidean three-space $\referenceE^3$ 
(which could in turn be viewed as a slice of Minkowski spacetime). Such 
{\em flat-space subtraction} assigns that portion of the auxiliary 
$\referenceE^3$ contained within $B$ the zero value of energy. Flat--space
subtraction does not modify the momentum densities $j_k$, since each of
these vanish for a two--surface drawn in an inertial $\referenceE^3$
hyperplane of Minkowski spacetime.

Finally, let us collect some overall results for referenced quasilocal 
energy and round spheres. Start with the general line-element for a 
spherically symmetric spacetime,
\be
ds^2 = -N^2 dt^2 + H^2 (dr + V^{r}dt)^{2} 
           + R^2 (d\theta^2 
+ \sin^2\theta d\phi^2)\, ,
\label{line_element}
\ee
where $N$, $H$, and the areal variable $R$ are functions of $t$ and $r$. 
Let $\Sigma$ be the 
interior of a  $t={\rm constant}$ slice with two--boundary $B$ specified 
by $r = {\rm constant}$. A straightforward calculation of the trace $k$ of the
extrinsic curvature $k_{ab}$ yields
\be
k = -\frac{2R'}{HR}\, ,
\ee
where the prime denotes $r$ partial differentiation. Now consider a round 
sphere with radius $R$ embedded in $\referenceE^{3}$. Such a sphere has an 
extrinsic curvature $(k^{0})_{ab}$ with trace
\be
k^0 = -\frac{2}{R}\, .
\ee
For the scenario at hand, the referenced energy density is then 
\be
\varepsilon = {1\over 4\pi} \biggl(\frac{1}{R} - \frac{R'}{HR}
              \biggr)\, ,
\label{surfacevarep}
\ee
from which we find
\be
E = R \bigl(1-R'/H\bigr)
\label{gensphereenergy}
\ee
for the total quasilocal energy (\ref{totalQLE}).

\subsection{Static solutions}
Let us assume the geometry is static, with $V^{r} = 0$ and $r = R$.
For a simple isentropic fluid with energy density $\rho(R)$ and pressure
$p(R)$, the Hamiltonian constraint $G^t_t = -8\pi \rho$ implies\cite{MTW}
\be
H = \biggl( 1 - {2m\over R}\biggr)^{-1/2} \, 
\ee
where
\be
m(R) = 4\pi\int_0^R d\bar R\, \bar R^2 \rho(\bar R) + M \, 
\label{fluid_mass}
\ee
The Schwarzschild black hole solution is obtained by 
choosing $\rho = p = 0$ and $m=M$, whereas a fluid star 
solution with $\rho \neq 0$ must have $M=0$ for
the geometry to be smooth at the origin. In each case, 
the energy is
\be
E = R\biggl[1-\biggl( 1-{2m(R) \over R}\biggr)^{1/2}\biggr]\, ,
\label{staticQLE}
\ee 
with $m(R)$ defined in Eq.~(\ref{fluid_mass}).
Observe that for a compact star or black hole, $E\to m(\infty)$ in the limit 
$R\to\infty$, which is precisely the {\sc adm} energy at infinity.\cite{ADM} 
Section VI further examines the relationship between the quasilocal energy 
(\ref{totalQLE}) and the {\sc adm} energy for more general asymptotically flat
spacetimes. We note that for the geometry at hand $K_{ij} = 0$, and 
it follows from the results of Section III that $\jmath_{\vdash} = \jmath_{a} 
= 0$; that is to say, all (unreferenced) quasilocal momentum densities vanish.

As discussed in Ref.~\cite{BY}, the Newtonian approximation for 
$E$ consists in assuming $m/R$ to be small, which yields
\be
E \approx m + {m^2\over 2R} \, .
\label{Newton_approx}
\ee
In this same approximation the first term, $m(R)$, is just the sum of the
matter energy density plus the Newtonian gravitational potential energy
associated with assembling the ball of fluid by bringing the individual 
particles together from infinity (see Ref.~\cite{MTW}, Box 23.1).  
The second term in Eq.~(\ref{Newton_approx}), namely
$m^2/ 2R$, is just {\it minus\/} the Newtonian gravitational potential energy
associated with building a spherical shell of radius $R$ and mass $m$, by
bringing the individual particles together from infinity. Thus, in the
Newtonian approximation, the energy $E$ has the natural interpretation as the
sum of the matter energy density plus the potential energy associated with
assembling the ball of fluid by bringing the particles together from the
boundary of radius $R$. In this sense, $E$ is the total energy of the system
contained within the boundary, reflecting precisely the energy needed to
create the particles, place them in the system, and arrange them in the final
configuration. Any energy that may be expended or gained in the process of  
bringing the particles to the boundary of the system, say, from infinity, is 
irrelevant.

\subsection{Boosted foliation of Schwarzschild}
To contrast the results obtained above, let us consider quasilocal energy-momentum 
for the Schwarzschild solution, but now with respect to a (radially) boosted 
foliation. For round spheres embedded in the preferred static time slices, we 
have found a referenced {\sc qle} (\ref{staticQLE}) that is a function of the 
areal radius $R$. The boosted foliation of Schwarzschild we now present has an
associated {\sc qle} which equals the mass parameter $M$ for any $R$ value. 
Such a foliation arises naturally when comparing the {\sc qle} considered here 
and the spinorial definition
\cite{DouganMason} of {\sc qle} given be Dougan and Mason.\cite{Lau1}
To obtain the new foliation, start with the Schwarzschild line-element written in
terms of the preferred 
static or curvature coordinates $(T,R)$,\cite{MTW}
\be
ds^{2} = - F dT^{2} + F^{-1} dR^{2} + R^{2}
(d\theta^2
+ \sin^2\theta d\phi^2)\, ,
\ee
where $F \equiv 1 - 2M/R$ (and $M$ is of course the mass parameter of the solution).
In terms of the new time coordinate${}^{14}$
\be
t \equiv T + M\log\left|\frac{1-F^2}{F^2}\right|\, ,
\label{newtime}
\ee
the line element is given by Eq.~(\ref{line_element}) with
\bea
r & = & R \, ,\label{lapse_shift} \eqnum{\ref{lapse_shift}a}\\
N = H^{-1} & = & {\textstyle \frac{1}{2}}(F + 1)\, , 
\eqnum{\ref{lapse_shift}b}\\
V^{r} & = & {\textstyle \frac{1}{4}}(F^{2} - 1)\, .
\eqnum{\ref{lapse_shift}c}
\addtocounter{equation}{1}
\eea
In Fig.~2 we plot some of the new foliation's level time slices in the right
exterior region of the Penrose diagram for Schwarzschild.
\begin{figure}[!htb]
\epsfxsize=4in
\centerline{\epsfbox{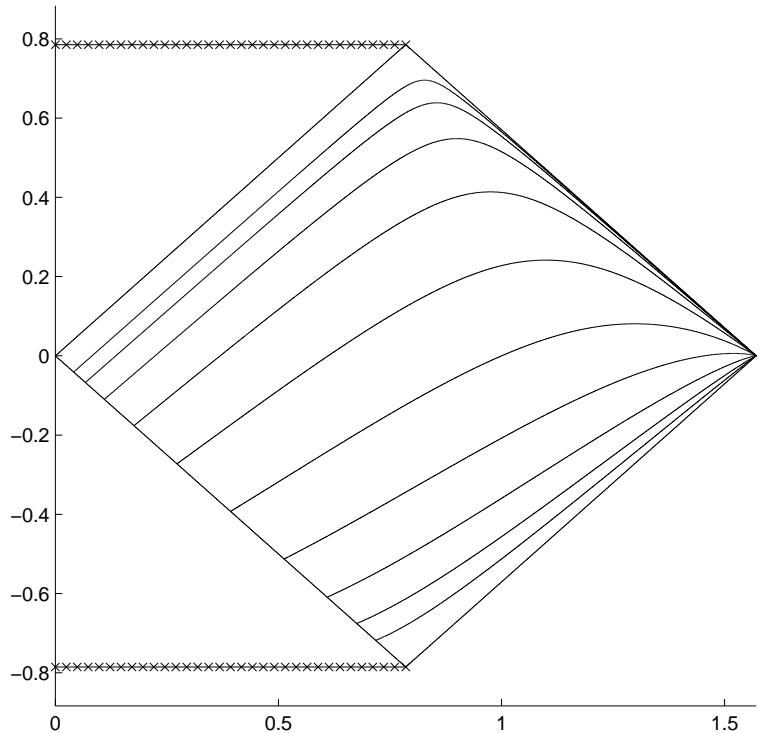}}
\caption{Penrose diagram for the exterior region of the 
Schwarzschild geometry.}
\end{figure}
The figure depicts the $t = {\rm constant}$ slices for the
time coordinate (\ref{newtime}) in the exterior region of the
Schwarzschild geometry. For the full construction and 
interpretation of the relevant Penrose diagram, see 
Ref.~\cite{HawkEll}. Here we define null coordinates
(see Ref.~\cite{MTW}, page 832)
\bea
\tilde{u} & = & - (R/2M-1)^{1/2} e^{R/4M} e^{-T/4M} 
\label{Kruskal} \eqnum{\ref{Kruskal}a} \\
\tilde{v} & = & (R/2M-1)^{1/2} e^{R/4M} e^{T/4M}\, ,
\eqnum{\ref{Kruskal}b}
\addtocounter{equation}{1}
\eea
where our $\tilde{u}$ and $\tilde{v}$ correspond respectively
to $w'(2M)^{-1/2}$ and $v'(2M)^{-1/2}$ of Ref.~\cite{HawkEll},
page 153.
Moreover, in Eqs.~(\ref{Kruskal}a,b) the curvature coordinate
$T = T(t,R)$, as determined by Eq.~(\ref{newtime}). Then the 
vertical and horizontal coordinates in the figure are defined as
\bea
\tau & = & {\textstyle \frac{1}{2}}
           \left[\tan^{-1}(\tilde{u})
       +   \tan^{-1}(\tilde{v})\right]
\label{Kruskal2} \eqnum{\ref{Kruskal2}a} \\
\rho & = & {\textstyle \frac{1}{2}}
           \left[\tan^{-1}(\tilde{u})
       -   \tan^{-1}(\tilde{v})\right]\, .
\eqnum{\ref{Kruskal2}b}
\addtocounter{equation}{1}
\eea
Starting from the bottom of the diagram, we have drawn level-time
slices corresponding to
$t = -6M, -4M, -2M, 0M, 2M, 4M, 6M, 8M, 10M, 12M$. While the 
coordinate $t$ provides a good foliation of the shown exterior
region, we note that these slices cross in the past dynamical 
region.

Now, with Eqs.~(\ref{lapse_shift}) and (\ref{Kishdot}) we find
$K^{\theta}_{\theta} = K^{\phi}_{\phi} = -M/R^{2}$ for the components 
of $K_{ij}$ needed to compute $\ell = - 8\pi\jmath_{\vdash}$, 
defined before in Eq.~(\ref{unbardensities}a). Hence, with these 
components and with Eqs.~(\ref{surfacevarep}) and (\ref{lapse_shift}b) 
we obtain
\bea
8\pi\varepsilon & = & \frac{2M}{R^2}
\label{kandlsbh} \eqnum{\ref{kandlsbh}a}\\
8\pi \jmath_{\vdash} & = & \frac{2M}{R^2}\, ,
\eqnum{\ref{kandlsbh}b}
\addtocounter{equation}{1}
\eea
again with $\kappa = 8\pi$.
It follows that both the total {\sc qle} (\ref{totalQLE}) and 
the total normal momentum $J_{\vdash}$, the proper surface integral
of $\jmath_{\vdash}$, both have value $M$ for a sphere of any radius 
$R$ embedded in one of the hypersurfaces depicted in Fig.~2. 

Now introduce a system boundary $\barT$ in the Schwarzschild exterior
region determined by fixation of the areal radius $R =$ const, in which 
case the level-$t$ spacelike slices will not meet $\barT$ orthogonally.
Observers at rest in the $t$ slices would see a changing
areal radius $R$, because with the $t$-foliation normal $u$ considered
as an operator $u[R] \neq 0$.
We can compute the (referenced) quasilocal energy surface density 
$\bar{\varepsilon}$ for observers comoving with $\barT$. Although the
energy density in Eq.~(\ref{surfacevarep}) was not barred, we may 
nevertheless use that equation along with $r = R$ and $H = F^{1/2}$ to find
\be
\bar{\varepsilon} = \frac{1}{4\pi R}\left(1
- \sqrt{1-\frac{2M}{R}}\right)\, ,
\label{barepSch}
\ee
the result for a radius-$R$ round sphere embedded in a level-$T$ 
static slice. Upon proper integration over $B$, Eq.~(\ref{barepSch}) gives
Eq.~(\ref{staticQLE}) (although here we would denote the {\sc qle} by 
$\bar{E}$). Note that because we have used $-2/R$ both for the 
implicit $k^0$ in Eq.~(\ref{kandlsbh}a) and for 
the implicit $\bar{k}{}^0$ in Eq.~(\ref{barepSch}),
the boost relations (\ref{epjboosts}) are no longer valid (see the
discussion in Sec.~V.A pertaining to boosting reference terms).
However, the point of this analysis --that our {\sc qle} (and for that 
matter our quasilocal momentum) is inherently observer dependent--
should be clear. There is in no strict sense {\em one} quasilocal energy 
expression for Schwarzschild.

\subsection{Isotropic Cosmology}

Consider the standard isotropic cosmological model with line element 
\be
ds^2 = - dt^2 + a^2(t)[ dr^2 + f^2(r) (d\theta^2 + \sin^2\theta d\phi^2)] \, .
\label{cosmosmetric}
\ee
Here, $f(r) = \sin r$, $r$, or $\sinh r$ depending on whether space has 
positive ($\chi = +1$), zero ($\chi = 0$), or negative ($\chi = -1$) 
curvature, respectively. The quasilocal energy within a sphere of coordinate radius $r$, 
embedded in a homogeneous surface $t = {\rm const}$, can be obtained from the general 
result (\ref{gensphereenergy}) for spherically symmetric spacetimes. Here, we 
compute the quasilocal energy within 
a sphere of fixed proper area $4\pi R^2$ (fixed areal radius $R$). That is, the 
history $\barT$ of the boundary $B$ is the surface $a(t)\, f(r) = R$, and the 
embedding hypersurface is chosen to be orthogonal to $\barT$. In the notation of 
Sec.~II, we are computing the quasilocal energy density $\bar{\varepsilon}$ as seen by 
the (barred) observers who are comoving with the system boundary $a(t)\, f(r) = R$. 
In the calculation below we omit the bars. 

The normal and tangent to the boundary $a(t)\, f(r) = R$ are given by
\bea
   n^\mu & = & (-\alpha f {\dot a},\  \alpha f'/a,\ 0,\ 0) \, ,
   \label{nandu} \eqnum{\ref{nandu}a}\\
   u^\mu & = & (\alpha f',\ -\alpha f {\dot a}/a,\ 0,\ 0) \, ,
   \eqnum{\ref{nandu}b}
   \addtocounter{equation}{1}
\eea
where $f' = df/dr$, ${\dot a} = da/dt$, and $\alpha \equiv (f'{}^2 - f^2 {\dot a}^2)^{-1/2}$. 
We note that, with the definition for $f(r)$, $\alpha$ becomes
\be
   \alpha = \left(1 - R^2(\chi + {\dot a}^2)/a^2 \right)^{-1/2} \, .
\ee
The mean curvature of the sphere is straightforward to compute, with the result 
\be 
   k = -\sigma^{\mu\nu} \nabla_\mu n_\nu = - \frac{2}{\alpha R} \, .
\ee 
Combining this with the flat-space subtraction term $k^0 = -2/R$, we find 
\be
   E = 8\pi R \left( 1 - \sqrt{1 - R^2(\chi + {\dot a}^2)/a^2} \right) \, 
\ee 
for the quasilocal energy. 

The result above can be written in a more suggestive form by invoking the Einstein equation
$(\chi + {\dot a}^2)/a^2 = 8\pi \rho/3$, where $\rho$ is the proper energy 
density of matter including a contribution $\Lambda/8\pi$ from the cosmological 
constant. The quasilocal energy then becomes 
\be 
   E = 8\pi R \left( 1 - \sqrt{1 - 8\pi \rho R^2/3} \right) \, .
\ee 
This result for quasilocal energy has the same form as that given in 
Eqs.~(\ref{fluid_mass},\ref{staticQLE}) 
for a static spherically symmetric solution, provided  $M = 0$ and $\rho = {\rm const}$ in  
the definition (\ref{fluid_mass}) for $m(r)$. Note, however, that the matter energy 
density $\rho$ is constant on the homogeneous slices $t = {\rm const}$, but is not 
constant on the hypersurfaces orthogonal to the system boundary $a(t)\, f(r) = R$. 

\subsection{Cylindrical Waves}

Finally, we turn to a non-spherically symmetric example, namely Einstein-Rosen 
cylindrical waves (see Ref.~\cite{Kuchar2} and references therein) determined 
by a line-element of the following form:
\be
ds^{2} = e^{\Gamma-\Phi}(-dt^2 + dr^2)
             + e^{-\Phi}r^{2}d\theta^2 
             + e^{\Phi}dz^2\, ,
\label{cylindricalelement}
\ee
where the coordinates $(t,r,\theta,z)$ range over their usual values.
For the line-element (\ref{cylindricalelement}) the vacuum Einstein 
equations become\cite{Kuchar2}
\be
   \partial^{2}_{t}\Phi
 - \partial^{2}_{r}\Phi
 - r^{-1}\partial_{r}\Phi = 0\, ,
\label{cylinderEqns} 
\ee
where the function $\Gamma$ is written in terms of the scalar field 
$\Phi$ as
\be
\Gamma(r,t) =
            \frac{1}{2}
            \int_{0}^{r} d\rho\,
            \rho\left[(\partial_{t}\Phi)^{2}
            + (\partial_\rho \Phi)^{2}\right]\, .
\label{Gammadefinition}
\ee
We assume smoothness at the origin, so that $\partial_{r}\Phi \rightarrow 0$ as
$r\rightarrow 0$.

Ashtekar and Varadarajan, among others, have examined the $3+1$ vacuum 
Einstein equations under the assumption of a spacelike Killing vector 
field.\cite{AshVar} They show via ``Killing vector field reduction'' 
that the system is equivalent to the $2+1$ Einstein equations coupled 
to ``fictitious'' matter. Computing the total energy of the 
$2+1$ system, they thereby obtain a $3+1$ formula for energy per unit 
length along the Killing direction. The class of Einstein-Rosen 
cylindrical waves determined by the line-element 
(\ref{cylindricalelement}) above --with two hypersurface orthogonal 
Killing vector fields-- constitute a special case of their analysis. 
In this particular case, it is $\Phi$ which plays the role of matter
from the $2+1$ viewpoint, and the expression they offer,
\be
E_{\rm AV} = \frac{1}{4}\left(1 - e^{-4C}\right)\, ,
\label{AshVarenergy}
\ee
is for unit-length energy along the $\partial/\partial z$ direction.
The $C$ in their expression is Thorne's so-called 
$C$-energy\cite{Thorne}, essentially just the total ``flat-space energy'' 
of the $\Phi$ scalar field,
\be
C = \frac{1}{16}\int^{\infty}_{0} dr\, r
    \left[(\partial_{t}\Phi)^{2}
            + (\partial_r \Phi)^{2}\right] \, .
\label{Cenergydefinition}
\ee

$E_{\rm AV}$ can be obtained directly from the integral (\ref{totalQLE}). 
Consider an infinitely long cylindrical two-surface $B$ determined by 
fixation of $r$ and $t$. $B$ is embedded in a $\Sigma$ hypersurface 
determined by fixation of $t$. Let us compute the total {\sc qle} 
(\ref{totalQLE}) for a section of the cylinder of
unit proper length in the $z$-direction. The mean curvature of the cylinder
reads
\be
k = - \frac{e^{(\Phi-\Gamma)/2}}{r}\, .
\ee
Now turn to the computation of the $k^{0}$ term. The $B$ metric 
functions are easily read off from the $B$ line-element,
\be
ds^2 = e^{\Phi}dz^2 + e^{-\Phi}r^{2}d\theta^{2}\, .
\ee
Note that over the two-surface $B$ the field $\Phi$ is a constant, and
from this we easily infer that for the auxiliary embedding of $B$ in 
$\referenceE^{3}$ the associated mean curvature of the cylinder side is
\be
k^{0} = - \frac{e^{\Phi/2}}{r}\, .
\ee
Hence, the energy (per unit proper length along the Killing direction
$\partial/\partial z$) obtained from ({\ref{totalQLE}) is
\be
E = \frac{1}{4} \left(1 - e^{-\Gamma/2}\right)\, .
\label{cylinderQLE}
\ee
Comparing the definition (\ref{Gammadefinition}) of $\Gamma(r,t)$ with the 
$C$-energy (\ref{Cenergydefinition}), we immediately see that the energy per 
unit proper length (\ref{cylinderQLE}) agrees with the Ashtekar-Varadarajan 
expression (\ref{AshVarenergy}) in the $r\rightarrow\infty$ limit.


\section{Energy-momentum at spatial infinity}

In this section we consider the quasilocal energy as applied to 
spacetimes that are asymptotically flat in spacelike 
directions\cite{ADM,ReggeTeitelboim,York5,BeigOMurchadha},${}^{15}$ 
and discuss its relationship with the standard treatment of energy at 
spatial infinity ({\sc spi}). 
We begin by recalling the key observation from Regge and 
Teitelboim,\cite{ReggeTeitelboim} namely, that for asymptotically flat 
spacetimes the gravitational Hamiltonian must have well defined functional 
derivatives and must preserve the boundary conditions on the fields. 
This observation leads to a two step prescription for building the 
gravitational Hamiltonian. (i) Starting with the ``base" 
Hamiltonian (\ref{basehamiltonian}) (the smeared Hamiltonian and 
momentum constraints), one adds boundary terms and imposes boundary 
conditions so that the resulting Hamiltonian has well defined 
functional derivatives. For the Hamiltonian to have 
well defined functional derivatives the boundary terms in the variation 
of the Hamiltonian must vanish under the assumed boundary conditions. (ii) 
One checks that the boundary conditions are preserved 
under evolution by the Hamiltonian. If they are not, then the boundary 
conditions and boundary terms must be modified. 

In essence, the canonical quasilocal formalism is an application 
of step (i) above to a manifold with boundary. 
That is, by working with the action in Hamiltonian form, we identify 
the appropriate boundary terms and boundary conditions that yield a 
Hamiltonian with well defined functional derivatives. The key difference 
between the quasilocal analysis and the asymptotically flat analysis is 
that in the quasilocal case we do not require, as in step (ii), that the
Hamiltonian should preserve the boundary conditions. The reason is 
the following. In the asymptotically flat case the spacelike hypersurfaces 
$\Sigma$ are Cauchy surfaces. Thus, the data on one slice $\Sigma$  
completely determines the future evolution of the system. For consistency
the evolved data must obey the boundary conditions, otherwise the Hamiltonian 
on future $\Sigma$ slices will not have well defined functional derivatives. 
On the other hand, in the quasilocal context, the surfaces $\Sigma$ are 
not Cauchy surfaces. They do not carry enough information to determine the 
future evolution of the system (the spacetime interior to $\partial{\cal M}$). 
Therefore, we see that in the quasilocal case step (ii) cannot, and should 
not, be taken.

Note that the boundary conditions for our quasilocal Hamiltonian 
(\ref{TheHamiltonian}) and for the Regge-Teitelboim Hamiltonian 
for asymptotically flat spacetimes\cite{ReggeTeitelboim} both include 
fixation of the boundary two-metric, lapse function, and shift vector. In the 
asymptotically flat case, the boundary is at {\sc spi} and the 
boundary values are determined through 
the specified asymptotic behaviors of the spatial metric, lapse, and shift. 
Although the boundary conditions for the two Hamiltonians are not specified 
in the same manner, they are at least ``compatible" with one another. Also note that 
with the choice of flat space subtraction, the quasilocal energy  
(obtained from our quasilocal Hamiltonian with $N = 1$ and $V^i = 0$ on the 
boundary) vanishes for flat spacetime. 
Likewise, the ADM energy (obtained from the Regge-Teitelboim 
Hamiltonian with $N \to 1$ and $V^i \to 0$ at infinity) vanishes for flat spacetime. 

To summarize, the quasilocal Hamiltonian and the Regge-Teitelboim Hamiltonian 
are both constructed from the base Hamiltonian by adding boundary terms and 
imposing ``compatible" boundary conditions, so that their functional derivatives 
are well defined. Furthermore, in both cases, the energy obtained 
from these Hamiltonians is referenced to zero for flat spacetime.
Given this close correspondence between the quasilocal and asymptotically flat 
analyses, it is not at all surprising that for asymptotically 
flat spacetimes the quasilocal energy agrees with the ADM 
energy in the limit that the spatial boundary $B$ is pushed to infinity. 
Although this result is not surprising, we find that a careful demonstration is 
nevertheless interesting and enlightening.${}^{16}$
The remainder of this section is devoted to this demonstration. 

\subsection{Asymptotic flatness}
For spacetimes that are asymptotically flat towards {\sc spi},
the $\Sigma$ gravitational initial-data set 
$(h_{ij}, K^{ij})$ obeys the following fall-off conditions:
\begin{eqnarray}
h_{ij} & = & f_{ij} + \alpha_{ij} \label{falloff}
\eqnum{\ref{falloff}a}\\
K^{ij} & = & O(r^{-2}) 
\eqnum{\ref{falloff}b}\, .
\addtocounter{equation}{1}
\end{eqnarray} 
Here $f_{ij}$ is a flat background three-metric and $\alpha_{ij}$ an 
$O(r^{-1})$ perturbation thereof. The radial coordinate 
$r \equiv (f_{ij} x^{i} x^{j})^{1/2}$ is defined in terms of 
coordinates $x^{k}$ that are Cartesian with respect to the $f_{ij}$ 
metric. In this section we assume that the two-surface $B$ is a 
level-$r$ surface, that is to say a round sphere in the $f_{ij}$ 
metric (although not quite round in $h_{ij}$). The scalar curvature 
of $B$ then obeys
\begin{equation}
      {\cal R} = 2r^{-2}
               + {}^{3}{\cal R} r^{-3}
               + O(r^{-3-\epsilon})
\label{BRicci}
\end{equation}
($\epsilon$ small and positive). We demand that $r$ is large enough such 
that ${\cal R} > 0$ everywhere on $B$ (see comments near the end of
Sec.~V.A pertaining to Weyl's problem). Adopting the usual distinction 
between ``little-oh'' and ``big-oh'' notation, we could write $o(r^{-3})$ 
instead of $O(r^{-3-\epsilon})$. The point being that $o(r^{-3})$ means
``decay faster than $r^{-3}$ order'' in this context. We call such a two-surface 
a {\em large sphere}. 

\subsection{Momentum at {\sc spi}}
Momentum is identified with the phase space generator that moves 
the fields along the orbits of a spatial vector field $\xi^i$. In 
moving the fields along $\xi^i$, the changes in the fields at 
a given point are given by minus the Lie derivative along $\xi^i$. 
Thus, the quasilocal momentum in the $\xi^i$ direction is defined 
as the on-shell value of the Hamiltonian (\ref{TheHamiltonian}) with 
vanishing lapse $N = 0$ 
and shift vector given by $V^i = - \xi^i$: 
\be 
   P_\xi \equiv H[0,-\xi] = \int_B d^2x \sqrt{\sigma}\, \xi^i \jmath_i \, .
\label{aflatmomentum}\ee 
Here, $\jmath_i = -2 n^k P_{ki}/\sqrt{h}$ is the quasilocal momentum 
density (\ref{jmath}). As discussed in Sec.~V.A, with Euclidean reference
there is no reference contribution to $\jmath_i$, because as it stands 
$\jmath_i$ would already vanish were 
$\Sigma$ an inertial $\referenceE^3$ hyperplane of Minkowski spacetime. In 
other words, the reference contribution $-2 (n^k P_{ki}/\sqrt{h})|^0$ to 
$\jmath_i$ vanishes for the Euclidean subtraction at hand. This is so 
because an inertial hyperplane of Minkowski spacetime has a vanishing
extrinsic curvature tensor.
Now consider the case in which space is 
asymptotically flat, and the boundary $B$ is pushed to infinity. 
If the vector $\xi$ to chosen to be an asymptotic translation, 
then we immediately find that the momentum $P_\xi$ above agrees 
with the ADM momentum at 
infinity.\cite{ADM,ReggeTeitelboim} If the vector $\xi$ is chosen 
to be an asymptotic rotation, then the momentum $P_\xi$ agrees 
with the angular momentum at infinity.\cite{ReggeTeitelboim} [Although
here appearing in the context of asymptotic flatness, note that the
integral (\ref{aflatmomentum}) is the most general expression for quasilocal
momentum in the $\xi^i$ direction, if we take $\jmath_i = - 2 n^k P_{ki}/\sqrt{h}
+ 2 (n^k P_{ki}/\sqrt{h})|^0$, now assuming that 
$-2 (n^k P_{ki}/\sqrt{h})|^0$ need not arise from Euclideain reference and
hence may be non-zero.]

\subsection{Energy at {\sc spi}}
Consider the quasilocal energy (\ref{totalQLE}), now with $B$ a large 
sphere in the sense described above.${}^{17}$
Let us show the equivalence between the {\sc qle} and the
{\sc adm} energy. Note that $B$ is a Riemannian submanifold of the 
Riemannian space $\Sigma$. Let us examine the geometry of this embedding 
in order to obtain an asymptotic expression for the integral (\ref{totalQLE}) 
in terms of the $\Sigma$ Riemann tensor.
Recall the Gauss-Codazzi-Mainardi constraints\cite{York} which 
relate the intrinsic geometry $\sigma_{ab}$ (or $\cal R$) and 
extrinsic geometry $k_{ab}$ (of $B$ in $\Sigma$) to the 
$\Sigma$ Riemann tensor. Among these is the
following:
\be
     {\cal R} - k{}^2 
     + k_{ab} k^{ab} 
     =
       \sigma^{ik}\sigma^{jl} R_{ijkl}\, .
\label{embed1}
\ee
As the $\Sigma$ Riemann tensor is completely determined by 
the Ricci tensor $R_{ij}$ in three dimensions, we may
replace the {\sc rhs} of Eq.~(\ref{embed1}) with Ricci
terms using the identity 
\be
\sigma^{ik}\sigma^{jl} R_{ijkl} = 
R - 2R_{\vdash\vdash}\, ,
\label{SigmaRicciRiemann}
\ee 
where $\vdash$ denotes contraction
with the $B$ normal $n^{k}$. Splitting $k_{ab} =: \frac{1}{2}
k\sigma^{ab} + (k^{\rm TF})^{ab}$ into its trace and trace-free
pieces, we rewrite Eq.~(\ref{embed1}) as
\be
     {\cal R} - {\textstyle \frac{1}{2}} k{}^2
     + (k^{\rm TF})_{ab} (k^{\rm TF}){}^{ab}
     = \sigma^{ik}\sigma^{jl} R_{ijkl}\, .
\label{embed2}
\ee
From this equation we may easily obtain the relevant leading 
order asymptotic relationship between $k$ and the $\Sigma$ 
Riemann tensor. Now, the fall-off of the $\Sigma$ metric 
$h_{ij}$ determines that
\be
\sigma^{ik}\sigma^{jl} R_{ijkl} = 
{}^{3}(\sigma^{ik}\sigma^{jl} R_{ijkl})r^{-3}
                              + o(r^{-3})\, ,
\ee
with the superscript $3$ meaning take the $O(r^{-3})$ piece of
that inside the parenthesis. Now we make the 
{\em Ans\"{a}tze}${}^{18}$
\bea
       k & = & 
       -2r^{-1}         + 
       {}^{2}k r^{-2} + o(r^{-2})
\label{k_abAnsatz} \eqnum{\ref{k_abAnsatz}a}\\
       (k^{\rm TF})_{ab} & = &  O(r^{-2})  
\eqnum{\ref{k_abAnsatz}b}
\addtocounter{equation}{1}
\eea
Using the {\em Ans\"{a}tze} (\ref{k_abAnsatz}) along with
the expansion (\ref{BRicci}) in Eq.~(\ref{embed2}),
we quickly find
\begin{equation}
         {}^{2}k =
           {\textstyle \frac{1}{2}}\,{}^{3}
           (\sigma^{ik}\sigma^{jl} R_{ijkl} - {\cal R})\, .
\end{equation}
Along a similar line, we can compute the asymptotic form of
$k^{0}$, also appearing in Eq.~(\ref{totalQLE}). Indeed,
striking the Riemann curvature term from the {\sc rhs} of
Eq.~(\ref{embed2}) and replacing all $k$'s with $k^{0}$'s,
we get a valid constraint for the auxiliary embedding of
$B$ in $\referenceE^{3}$ [see Eq.~(\ref{Weylproblem}a)]. An 
asymptotic analysis of the resulting equations shows that 
\be
         {}^{2}k^{0} =
           - {\textstyle \frac{1}{2}}\,{}^{3}{\cal R}\, .
\ee
It follows that the quasilocal energy surface density
$\varepsilon = (k-k^{0})/(8\pi)$ obeys
\be
8\pi \varepsilon =
           {\textstyle \frac{1}{2}}\,{}^{3}
           (\sigma^{ik}\sigma^{jl} R_{ijkl}) + o(r^{-2})
\label{asymptoticEp}
\ee
in the asymptotic setting described here.

Eq.~(\ref{asymptoticEp}) shows that the {\sc spi} definition 
of energy determined by the canonical quasilocal energy
(\ref{totalQLE}) is equivalent to the (rather symbolic) integral
\be
       E_{\infty} =
       \frac{1}{16\pi} \oint_{\infty}d^{2}x 
       \sqrt{\sigma}\, r \sigma^{ik}\sigma^{jl} R_{ijkl},
       \label{SigmaRiemannintegral}
\ee
where we use $\oint_{\infty}$ as shorthand for 
$\lim_{r\rightarrow\infty}\int_{B}$ so that 
$\sqrt{\sigma}$ tends to $r^2\sin\theta$.
We can rewrite this integral in a familiar form. Recall 
that $\Re_{ijkl} = R_{ijkl}\,+$ terms quadratic in the 
extrinsic tensor $K_{ij}$ [and hence $O(r^{-4})$ terms],
where $\Re_{ijkl}$ is the $\Sigma$ projection of the 
spacetime Riemann tensor $\Re_{\mu\nu\lambda\kappa}$ (the 
same as the Weyl tensor $C_{\mu\nu\lambda\kappa}$ near {\sc spi}).
Moreover, we may use $r \sim \sqrt{A/(4\pi)}$ where $A$ is 
the area of $B$. Therefore, the integral above agrees 
asymptotically with what Hayward\cite{SHayward} calls the 
Ashtekar-Hansen expression${}^{19}$
\begin{equation}
E_{\rm AH} =
       \frac{1}{8\pi}\sqrt{\frac{A}{16\pi}}
       \int_{B}d^{2}x
       \sqrt{\sigma}\,\sigma^{\mu\lambda}\sigma^{\nu\kappa} 
       C_{\mu\nu\lambda\kappa}\, ,
\label{AHintegral}
\end{equation}
which is quite similar to Hayward's mass Eq.~(\ref{Haywardmass}).
The Ashtekar-Hasen expression agrees asymptotically with
\be
E_{\rm ADM} = \frac{1}{16\pi}
\oint_{\infty} d^{2}x \sqrt{\sigma}\,n^{k}
\left(\partial^{j}\alpha_{kj} - \partial_{k} \alpha^{j}{}_{j}
\right)\, ,
\label{ADMintegral}
\ee
the familiar {\sc adm} energy. We can replace $\sqrt{\sigma}\,n^{k}$ with 
$r x^{k}\sin\theta$ in this integral.

Although apparently a result known for some time\cite{Geroch}, let us briefly 
argue why the integrals (\ref{SigmaRiemannintegral}) 
and (\ref{ADMintegral}) agree in the {\sc spi} limit considered
here. A more detail account of this issue and others concerning 
surface integrals at {\sc spi} will appear elsewhere.\cite{BBLP} 
First, replace the Riemann curvature term in the integral 
(\ref{SigmaRiemannintegral}) in favor of Ricci terms using the 
identity (\ref{SigmaRicciRiemann}). Next, via standard 
techniques, obtain the identity
\be
       {}^{3}\!R_{ij} = {}^{3}\!\left[\partial^{p} \partial_{(i}
\alpha_{j)p}
     - {\textstyle \frac{1}{2}}\partial^{k}\partial_{k}\alpha_{ij}
     - {\textstyle \frac{1}{2}}\partial_{i}\partial_{j}
       \alpha^{p}{}_{p}\right]\, , \label{3piece}
\ee
where the $\partial_{k}$'s are Cartesian partial derivatives.
Note that one is taking the $O(r^{-3})$ piece of the
{\sc rhs} of the above equation. Finally, use this last identity 
along with the useful fact that $\partial_{r} \alpha_{ij} = 
- r^{-1} \alpha_{ij} + o(r^{-2})$, to show that --at leading order-- 
the integrands in (\ref{SigmaRiemannintegral}) and (\ref{ADMintegral}) 
differ by a pure divergence on the unit sphere.

\section{Acknowledgments}
This work was conducted over a
eight year period of time, 
beginning in 1993. During this time, 
SRL visited the Raman
Research Institute on two separate occasions, and 
he doubly thanks that institute, and in particular 
J.~Samuel, for hospitality. JDB would like to thank the 
Department of Mathematics, North Carolina State University, 
for support during the early stages of this work. 
We also thank both J.~Jezierski for discussions and for 
bringing Ref.~\cite{Kijowski} to our attention and A.~N.~Petrov 
for several email communications on the topic of energy at spatial 
infinity. This work was supported in part by National Science 
Foundation grants PHY-0070892 to North Carolina State 
University, and PHY-9972582 to the University of North Carolina. 

\noindent
{\em Note Added:} Recently, Anco and Tung have also carried out 
a careful systematic analysis of gravitational boundary conditions
for a quasilocal spacetime region, one based on a general Noether
charge formulation.\cite{AncoTung}
    
\appendix
\section{Kinematics of a double foliation}
Consider a double foliation of spacetime into $t={\rm const}$ 
surfaces and $s={\rm const}$ surfaces, where $t={\rm const}$ 
is spacelike with leaves $\Sigma$
and $s={\rm const}$ is timelike with one of its leaves $\barT$. Let 
\bea 
u_\mu &=& -N\nabla_\mu t \ ,\label{normals}
\eqnum{\ref{normals}a}\\
\barn_\mu &=& \barM \nabla_\mu s \ ,\eqnum{\ref{normals}b}
\addtocounter{equation}{1}
\eea
denote the unit-normal covectors of the $t={\rm const}$ and 
$s={\rm const}$ surfaces, respectively. Note that $N$ and 
$\barM$ serve to normalize these vectors, so $u\cdot u = -1$
and $\barn\cdot \barn = +1$. The induced (projection) 
metrics for the $t$ and $s$ foliations are 
\bea 
h_{\mu\nu} &=& g_{\mu\nu} 
+ u_\mu u_\nu \ ,\label{indmetrics}\eqnum{\ref{indmetrics}a}\\
\bargam_{\mu\nu} &=& g_{\mu\nu} 
- \barn_\mu \barn_\nu \ ,\eqnum{\ref{indmetrics}b}
\addtocounter{equation}{1}
\eea 
respectively. Also define the new unit vector fields 
\bea 
n_\mu &=& M D_\mu s = M h_\mu^\nu \nabla_\nu s 
= (M/\barM)h_\mu^\nu \barn_\nu 
\ ,\label{newnormals}\eqnum{\ref{newnormals}a}\\
\baru_\mu &=& -\barN \D_\mu t 
= -\barN \bargam_\mu^\nu \nabla_\nu t 
= (\barN/N)\bargam_\mu^\nu u_\nu \ ,\eqnum{\ref{newnormals}b}
\addtocounter{equation}{1}
\eea 
where $D_\mu$ is the covariant derivative on $t={\rm const}$, 
and $\D_\mu$ is the covariant derivative on $s={\rm const}$. 
$M$ and $\barN$ are chosen so that $n\cdot n = +1$ and 
$\baru\cdot\baru = -1$. The covectors $n_\mu$, $\baru_\mu$ lie 
in the $t={\rm const}$, $s={\rm const}$ surfaces, respectively, 
and are orthogonal to the two--surfaces formed by the 
intersections of the double foliation. In general $n_\mu$ 
and $\baru_\mu$ are not three--surface orthogonal. For example, whereas 
$\baru^{\mu}$ is indeed two--surface orthogonal as a vector field within 
a {\em single} $s = {\rm const}$ three--surface (as the future--pointing 
normal to the $B$ foliation), $\baru^{\mu}$ need not define 
a foliation of $\M$ into spacelike three--surfaces. 
The spacetime representation of the $B$ two--metric may now 
be expressed in two ways:
\be
\sigma_{\mu\nu} = g_{\mu\nu} + u_\mu u_\nu - n_\mu n_\nu = 
g_{\mu\nu} + \baru_\mu \baru_\nu - \barn_\mu \barn_\nu {\,} .
\label{twometric}
\ee 

Now define a time flow vector field $t^\mu$ along the $s={\rm const}$ 
surfaces by the conditions 
\be 
t^\mu\nabla_\mu t = 1 \ ,\qquad t^\mu\nabla_\mu s = 0 \ .
\label{flow}\eqnum{\ref{flow}a}
\ee
Likewise, fix a space flow vector field $s^\mu$ along the $t={\rm const}$ 
surfaces by the conditions 
\be 
s^\mu\nabla_\mu s = 1 \ ,\qquad s^\mu\nabla_\mu t = 0 \ .\eqnum{\ref{flow}b}
\addtocounter{equation}{1}
\ee
Then the shift vectors for the double foliation are given by 
\bea 
V^\mu &=& h^\mu_\nu t^\nu \ ,\label{shifts}\eqnum{\ref{shifts}a}\\
\barW^\mu &=& \bargam^\mu_\nu s^\nu \ .\eqnum{\ref{shifts}b}
\addtocounter{equation}{1}\eea
The time and space flow vector fields can also be written as 
\bea 
t^\mu &=& Nu^\mu + V^\mu \ ,
\label{otherflows}
\eqnum{\ref{otherflows}a}\\
s^\mu &=& \barM \barn^\mu + \barW^\mu \ , 
\eqnum{\ref{otherflows}b}
\addtocounter{equation}{1}\eea
respectively. 

Using the expressions (\ref{normals}b) for the normal $\barn_\mu$ and 
(\ref{otherflows}a) for the time flow vector field $t^\mu$, we find that
$u^\mu\barn_\mu  = (\barM/N)(t^\mu - V^\mu)\nabla_\mu s = -(\barM/M)v$, where 
the proper ``radial" velocity is defined by $v = V\cdot n/N$. In deriving this result, 
Eqs.~(\ref{newnormals}a) and (\ref{flow}a) were used.  
A similar calculation  using the expressions (\ref{normals}a) and 
(\ref{otherflows}b) for the  normal $u_\mu$ and the 
space flow vector field $s^\mu$ yields $\barn^\mu
u_\mu = -(N/\barM)(s^\mu -\barW^\mu)\nabla_\mu t  = -(N/\barN)(\barW\cdot\baru/\barM)$. 
Putting these  results together, we have (using a ``$\cdot$'' to denote spacetime
inner product)
\be 
u\cdot\barn = -\frac{\barM}{M}v 
= -\frac{N}{\barN}\left(\frac{\barW\cdot\baru}{\barM}\right) \ .
\label{udotnbar1}
\ee
The normalization condition for $n_\mu$ implies 
\bea
M &=& (D_\mu s\, g^{\mu\nu}\, D_\nu s)^{-1/2} 
= (\nabla_\mu s\, h^{\mu\nu}\, \nabla_\nu s)^{-1/2} \nonumber\\
&=& \barM(\barn_\mu h^{\mu\nu} \barn_\nu)^{-1/2} 
= \barM(1 + (u\cdot\barn)^2 )^{-1/2}  \nonumber\\
&=& \barM (1 + (\barM v/M)^2 )^{-1/2} \ .\label{Mequation}
\eea 
Solving for $M$, we find  $M = \barM/\gamma$ where $\gamma = (1-v^2)^{-1/2}$. This 
implies, from Eq.~(\ref{udotnbar1}), that
$u\cdot\barn = -\gamma v$. A calculation similar to  the one in Eq.~(\ref{Mequation}) 
for the normalization
condition on $\baru_\mu$ gives $\barN = N/\gamma$.  Now Eq.~(\ref{udotnbar1}) shows that 
$(\barW\cdot\baru /\barM)= v$. To summarize the results thus far, we have 
\bea 
v &=&\frac{V\cdot n}{N} = \frac{\barW\cdot\baru}{\barM} \ ,
\label{vandgamma}\eqnum{\ref{vandgamma}a}\\
\gamma &=& \frac{N}{\barN} = \frac{\barM}{M} \ ,\eqnum{\ref{vandgamma}b}
\addtocounter{equation}{1}
\eea 
where $\gamma = (1-v^2)^{-1/2}$.

Our next task is to express the barred unit vectors in terms of the unbarred 
unit vectors. From the definition (\ref{newnormals}b) 
of $\baru_\mu$,  we have 
\bea 
\baru_\mu &=& (\barN/N)\bargam_\mu^\nu u_\nu \nonumber\\
&=& (\barN/N) (u_\mu - (u\cdot\barn)\barn_\mu) \nonumber\\
&=& (1/\gamma) (u_\mu +\gamma v \barn_\mu) \ ,\label{baru}
\eea
with a similar calculation showing that 
\be 
n_\mu = (1/\gamma)(\barn_\mu -\gamma v u_\mu) \ .\label{enn} 
\ee 
Putting these together, we find 
\bea 
\barn_\mu &=& \gamma n_\mu + \gamma v u_\mu \ ,
\label{appenAbarrelations}\eqnum{\ref{appenAbarrelations}a}\\
\baru_\mu &=& \gamma u_\mu + \gamma v n_\mu \ .
\eqnum{\ref{appenAbarrelations}b} 
\addtocounter{equation}{1}
\eea
Equivalently, we obtain 
\bea 
n_\mu &=& \gamma \barn_\mu - \gamma v \baru_\mu \ ,
\label{relations}\eqnum{\ref{relations}a}\\
u_\mu &=& \gamma \baru_\mu - \gamma v \barn_\mu \ ,
\eqnum{\ref{relations}b} 
\addtocounter{equation}{1}
\eea 
by inverting Eqs.~(\ref{appenAbarrelations}). 

We now derive two useful expressions, one for  $\baru^\mu$ in terms of 
$t^\mu$ and the other for $n^\mu$ in terms of $s^\mu$. Begin with the definition 
(\ref{otherflows}a) 
for the time flow vector field $t^\mu$ and write the shift vector as 
$V^\mu = h^\mu_\nu V^\nu = \sigma^\mu_\nu V^\nu + N^\mu (V\cdot n)$. Using 
the formulas (\ref{vandgamma}) and (\ref{barrelations}a) we find 
\be 
\barN \baru^\mu = t^\mu - \sigma^\mu_\nu V^\nu \ . 
\ee 
A similar calculation starting from the definition (\ref{otherflows}b) for the space 
flow vector field $s^\mu$ yields 
\be 
M n^\mu = s^\mu - \sigma^\mu_\nu \barW^\nu \ , 
\ee 
where equations (\ref{vandgamma}) and (\ref{relations}a) are used. 

The foliation of spacetime into $t={\rm const}$ surfaces is pictured  
in Fig.~3. 
\begin{figure}[!htb]    
\epsfbox{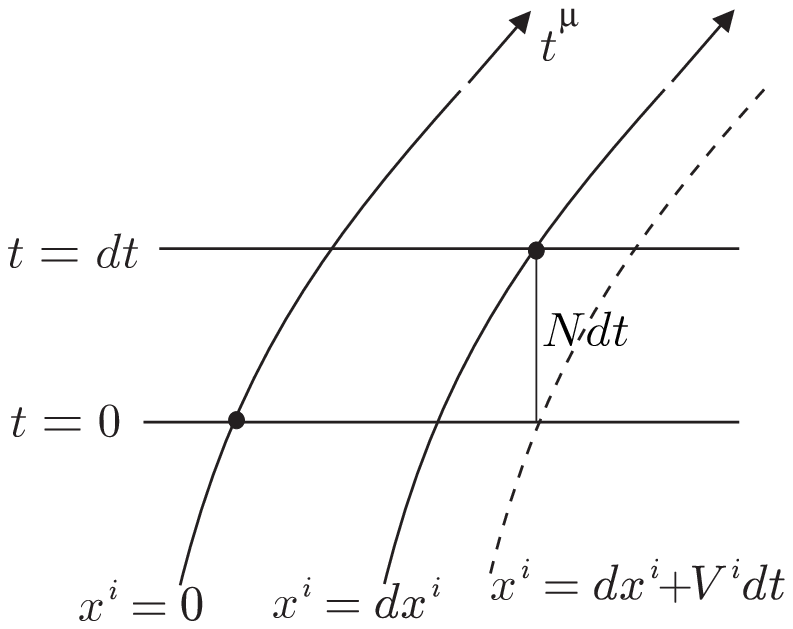}    
\caption{}       
\end{figure} 
The proper time between $t={\rm const}$ slices, measured orthogonal to the 
slices, is $N\,dt$. In the diagram the shift vector
$V^\mu$ points to the right, along the direction of increasing $x^i$, so the component $V^i$ is 
positive. The proper distance between the 
heavy dots is 
\be 
ds^2 = -N^2 dt^2 + h_{ij}(dx^i + V^i dt)(dx^j + V^j dt) \ , 
\ee 
where $h_{ij}$ is the metric on $t={\rm const}$. 

The foliation of spacetime into $s={\rm const}$ surfaces is pictured in 
Fig.~4.
\begin{figure}[!htb]   
\epsfbox{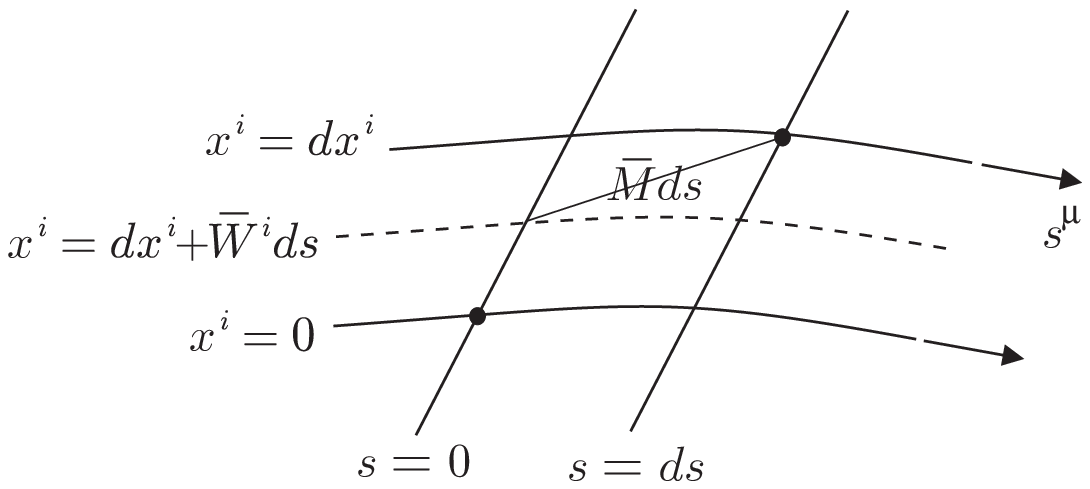}      
\caption{}    
\end{figure} 
The proper distance between $s={\rm const}$ slices, measured orthogonal 
to the slices, is $\barM\,ds$. In the diagram the shift vector
$\barW^\mu$ points to the past, along the direction  of decreasing $x^i$, so the component $\barW^i$
is negative. The proper distance  between the heavy dots is 
\be 
ds^2 = \barM^2 ds^2 + \bargam_{ij}(dx^i + \barW^i ds)(dx^j + \barW^j ds) \ , 
\ee 
where $\bargam_{ij}$ is the metric on $s={\rm const}$. 

\section{Extrinsic Curvature ``splittings''}
Let us now derive the splitting expression (\ref{barTsplit}) as well as
easier splitting (\ref{Ksplit}) for $K_{ij}$. Recall that the $\barT$ extrinsic
curvature is defined by 
$\barTheta_{\mu\nu} \equiv -\bargam^\alpha_\mu\nabla_\alpha \barn_\nu$. 
With the identity 
\be
\nabla_{\mu} u_{\nu} = - K_{\mu\nu}
- u_{\mu} a_{\nu} \ ,\label{B.3}
\ee
and the definitions $v\gamma = - u \cdot \barn$ and $\gamma = (1-v^2)^{-1/2}$, 
it is straightforward to verify that
\be
\nabla_{\mu} \barn_{\nu} = \gamma \,\nabla_{\mu} n_{\nu} 
- \gamma v\, K_{\mu\nu} - \gamma v\, u_{\mu}  a_{\nu} + 
\gamma^{2}\,\baru_{\nu}\,\nabla_{\mu} v\ . \label{B.4} 
\ee 
From this result it follows that the projection of $\barTheta_{\mu\nu}$ into $B$ is 
$\sigma^\alpha_\mu \sigma^\beta_\nu \barTheta_{\alpha\beta} = \gamma k_{\mu\nu} 
+ \gamma v \sigma^\alpha_\mu \sigma^\beta_\nu K_{\alpha\beta}$, where 
$k_{\mu\nu} = - \sigma^\alpha_\mu D_\alpha n_\nu$ is the extrinsic 
curvature of $B$ embedded in $t={\rm const}$. 
The identity (\ref{B.3}) and the definition (\ref{kandl}b) show that the 
term $\sigma^\alpha_\mu \sigma^\beta_\nu K_{\alpha\beta}$ equals the 
extrinsic curvature $\ell_{\mu\nu}$ of $B$ embedded in a surface whose 
normal at $B$ coincides with  $n^\mu$. Thus, we find 
\be 
\sigma^\alpha_\mu \sigma^\beta_\nu \barTheta_{\alpha\beta} 
= \gamma k_{\mu\nu} + \gamma v \ell_{\mu\nu} \ .\label{B.5} 
\ee
Next, we note that 
\bea 
\baru^\mu \baru^\nu \barTheta_{\mu\nu} 
&=& - \baru^\mu \baru^\nu \nabla_\mu \barn_\nu \nonumber \\
&=& \barn\cdot \bara \ ,\label{B.6} 
\eea 
where we have used the Leibniz rule $\baru^\nu\nabla_\mu \barn_\nu = 
\nabla_\mu(\baru^\nu\barn_\nu) - \barn_\nu\nabla_\mu \baru^\nu 
= - \barn_\nu\nabla_\mu\baru^\nu $ and the definition 
$a^\nu = \baru^\mu \nabla_\mu \baru^\nu$ of the acceleration of $\baru^\nu$. 
Finally, we compute 
\bea
\sigma^\alpha_\mu \baru^\beta \barTheta_{\alpha\beta} 
& = & -\sigma^{\alpha}_{\mu}  \baru^{\beta}  \nabla_{\alpha} \barn_{\beta} 
\nonumber \\ 
& = & -\sigma^{\alpha}_{\mu} 
\baru^{\beta}\left(\gamma\, \nabla_{\alpha} n_{\beta} 
- \gamma v\, K_{\alpha\beta} - \gamma v\, u_{\alpha} a_{\beta} 
+ \gamma^{2}\,\baru_{\beta} \nabla_{\alpha} v\right) \nonumber \\
& = &  -\gamma^{2}\, \sigma^{\alpha}_{\mu}  u^{\beta} \nabla_{\alpha} 
n_{\beta} + \gamma^2 v^{2}  \sigma^{\alpha}_{\mu}  n^{\beta}  
K_{\alpha\beta}  + \gamma^{2}\, 
\sigma^{\alpha}_{\mu} \nabla_{\alpha} v \ ,\label{B.7} 
\eea
where Eqs.~(\ref{barrelations}a) and (\ref{B.4}) have been used. 
Application of the Leibniz rule on the 
first term  on the right--hand side above yields $\sigma^\alpha_\mu u^\beta 
\nabla_\alpha n_\beta = - \sigma^\alpha_\mu n_\beta \nabla_\alpha u^\beta 
= \sigma^\alpha_\mu n^\beta K_{\alpha\beta}$. Collecting terms, we find 
\be
\sigma^\alpha_\mu \baru^\beta \barTheta_{\alpha\beta}  =
-\sigma^{\alpha}_{\mu} n^{\beta}  K_{\alpha\beta} 
+ \gamma^{2} \sigma^{\alpha}_{\mu} \nabla_{\alpha} v\ .\label{B.8} 
\ee
The results (\ref{B.5}), (\ref{B.6}), and (\ref{B.8}) then give Eq.~(\ref{barTsplit}).

The expression (\ref{barTsplit}) for $\barTheta_{\mu\nu}$ still retains 
reference to barred quantities through the  appearance of $\baru^\mu$ 
and $\barn \cdot \bara$. Definitions (\ref{barrelations}) can be  used to 
eliminate $\baru^\mu$ in favor of $u^\mu$ and $n^\mu$. The acceleration 
$\barn \cdot \bara$ can be re-expressed with the help of Eq.~(\ref{B.4}): 
\bea 
\barn \cdot \bara &=& \barn_\nu \baru^\mu \nabla_\mu \baru^\nu \nonumber\\
&=& - \baru^\mu \baru^\nu \nabla_\mu \barn_\nu \nonumber\\
&=& - \baru^\mu \baru^\nu \left(\gamma 
\nabla_{\mu} n_{\nu} 
- \gamma v\,  K_{\mu\nu} - \gamma v\, u_{\mu}  a_{\nu} 
+ \gamma^{2}\baru_{\nu}\,\nabla_{\mu} v\right) \nonumber \\
&=& - \gamma^{2} \baru^{\mu}  u^{\nu} 
\nabla_{\mu} n_{\nu} 
+ \gamma^{3} v^3 n^{\mu} n^{\nu} K_{\mu\nu} 
- \gamma^{3} v^{2}  n\cdot a  
+ \gamma^{2} \baru^{\mu}\nabla_{\mu} v\ .\label{B.10} 
\eea
The first term on the 
right--hand side of the last line can be re-expressed as 
\bea
\baru^{\mu}\, u^{\nu} \nabla_{\mu} n_{\nu} & = &
\baru^{\mu}  n^{\nu}\left( K_{\mu\nu} + u_{\mu}
a_{\nu}\right) \nonumber \\
& = &  \gamma  v\, n^{\mu} n^{\nu}  K_{\mu\nu} 
- \gamma \, n\cdot a \ , \label{B.11} 
\eea
which leads to 
\be
\barn\cdot \bara  = - \gamma v\, 
n^{\mu} n^{\nu}  K_{\mu\nu} + \gamma\, n\cdot  a 
+ \gamma^{2}  \baru^{\mu} \nabla_{\mu}v \ .\label{B.12} 
\ee
The extrinsic curvature term $n^\mu n^\nu K_{\mu\nu}$ can be rewritten 
using the identity (\ref{B.3}) followed by an integration by parts; this leads to 
$n^\mu n^\nu K_{\mu\nu} = u\cdot b$, where $b^\mu = n^\nu \nabla_\nu n^\mu$.  
Collecting results, we find 
\be 
\barn \cdot \bara = \gamma\, n\cdot a - \gamma v\, u\cdot b 
+ \baru\cdot\nabla\theta \ .\label{B.13} 
\ee 
This is Eq.~(\ref{accboost}a) from the main text. 

The derivation of the splitting (\ref{Ksplit}) amounts to simply projecting $K_{ij}$
into various pieces normal and tangential to $B$. We leave this to the reader. Let
us however sketch the derivation of Eq.~(\ref{accboost}b). Using the boost relations
(\ref{barrelations}) and the identity (\ref{B.3}), we find
\be 
\nabla_\alpha \baru_\nu = \gamma v \nabla_\alpha n_\nu - \gamma K_{\alpha\nu} 
- \gamma u_\alpha a_\nu + \gamma^2 \barn_\nu\nabla_\alpha v \ , \label{B.16} 
\ee 
from which the quantity $\barn\cdot \barb = - \barn^\mu \barn^\nu \nabla_\mu \baru_\nu$ 
can be expressed in terms of unbarred  quantities [with 
result (\ref{accboost}b)]. 

\section{Velocity parameter $\theta$ and its time rate of change}
Let us turn to the interpretation of $\dot\theta$ defined in 
Eq.~(\ref{roctheta}). We have defined $\dot\theta$ as such in order
that $\theta$ is the velocity parameter. We may verify the correctness
of our assignment by considering boosts in flat
spacetime. Let
\bea
T &=& x\sinh t \ ,
\label{A1.1}\eqnum{\ref{A1.1}a}\\
X &=& x\cosh t \ ,\eqnum{\ref{A1.1}b}
\addtocounter{equation}{1}
\eea
where $X$, $T$ are Minkowski coordinates and $x$, $t$ are Rindler
coordinates. See Fig.~5.
\begin{figure}[!htb]   
\epsfbox{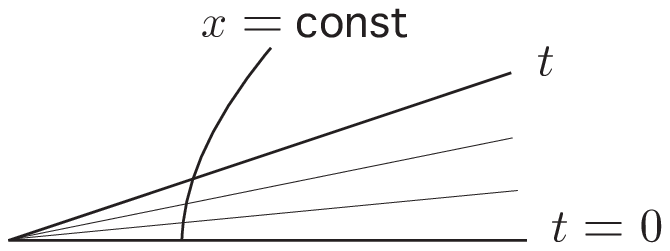}         
\caption{}  
\end{figure} 
The metric is $ds^2 = -dT^2 + dX^2 = -x^2dt^2 + dx^2$.
The four--velocity $U(t)$ along
$x={\rm const}$ at Rindler time $t$ is
$U(t) = (1/x)(\partial/\partial t) = \cosh t (\partial/\partial T)
+ \sinh t (\partial/\partial X)$.
The relativistic gamma factor between the four--velocities at $t=0$ and
$t$ is${}^{20}$
$\gamma = -U(t)\cdot U(0) = \cosh t$.
Now let $\tau = xt$ denote the proper time along $x={\rm const}$
between
$t=0$ and $t$. The rate of change of $\tau$
with respect to proper distance along $t={\rm const}$ is
$d\tau /dx = t = \cosh^{-1}\gamma$. This also can be expressed in
terms of the gradient of the lapse function: $d\tau/dx = \int_0^t
(n^i\partial_i\eta)$
Thus, we find
$\gamma = \cosh \int_0^t (n^i\partial_i\eta)dt$. From the definition of
the
velocity parameter, we have $\gamma = \cosh \theta$ so that
\be
\theta(t) = \int_0^t(n^i\partial_i\eta)dt \, .
\ee
It follows that $n^i\partial_i \eta = {\dot\theta}$, from which we obtain
Eq.~(\ref{roctheta}). Ref.~\cite{ArrestedHistory} discusses these results
in terms of degenerate foliations.

\newpage
\begin{center}
\large
{\bf FOOTNOTES}
\end{center}
\vskip 15pt
\noindent
1. This is not an exhaustive list.

\vskip 3mm \noindent
2. See also the related work
\cite{Freud} by Freud.

\vskip 3mm \noindent
3. For a comparison of the
Trace-K action with the tetrad action of Goldberg, see   
Ref.~\cite{Lau2}.

\vskip 3mm \noindent
4. With the
Trace-K action, as with any action, we are free to subtract a
function of the fixed boundary data. In some cases, it is   
convenient to fix this ambiguity by introducing a background
space or spacetime. However, this is not a required feature of the
{\sc cqf}. See Sec.~V.A for further discussion.

\vskip 3mm \noindent
5. Thus
both $\Sigma'$ and $\Sigma''$ are, as the
notation suggests, leaves of the $\Sigma$ foliation. We have
investigated the more general case in which $\Sigma'$ and $\Sigma''$
are not leaves of $\Sigma$, in which case some
portions of the boundaries of some $\Sigma$ leaves lie
in the boundary elements $\Sigma''$ or $\Sigma'$. However,
at present we find no compelling reason to allow for such generality.

\vskip 3mm \noindent
6. By introducing a partial foliation of
$\M$
that includes $\barT$ as one leaf, we can define ${\bar u}^\mu$ and  
$n^\mu$ as unit vector fields in a spacetime neighborhood of $\barT$.
Note, however, that as spacetime vector fields, ${\bar u}^\mu$ and $n^\mu$
are not in general hypersurface orthogonal. See Appendix A.

\vskip 3mm \noindent
7. A few words
concerning
terminology are in order. When we apply the {\em variational principle},
we
vary the action functional among all histories that
satisfy certain specified boundary conditions. The histories that
extremize   
the action under such a variation are, by definition, the classical
histories.
On the other hand, a {\em Hamilton--Jacobi variation}
({\sc hj} variation) of the action is a variation
among classical histories with different boundary values.

\vskip 3mm \noindent
8. Or, indeed, useful when examining any
variation involving a Ricci scalar curvature (we have in
mind the Ricci--scalar term
present in the initial--value Hamiltonian constraint).

\vskip 3mm \noindent
9. For $u^{\mu}$, of course,
$\epsilon = - 1$, but we shall
keep track of $\epsilon$ in order to ensure that the lemma
also holds for $\bar{n}^{\mu}$.

\vskip 3mm \noindent
10. However, the reader should resist the temptation
to identify $\baru^{\mu}$ with the future--pointing normal of $\barSigma$,
as in
our formalism $\baru^\mu$ need not be three--surface orthogonal. It is the
case
that the $\barSigma$ normal agrees with $\baru^\mu$ on the two--surface  
$B$ where $\barSigma$ and $\barT$ intersect. In fact, this is all that we  
require of $\bar\Sigma$, so in effect $\barSigma$ represents an
equivalence
class of three--slices determined by this condition on $B$.

\vskip 3mm \noindent
11. In that paper, the bars were omitted.

\vskip 3mm \noindent
12. The most effective way to proceed is to use the
result for the variation of the Hamiltonian, derived below.

\vskip 3mm \noindent
13. Nevertheless, let us present some details here. Typically, the
subtraction term will arise as a sum of functionals (smearing integrals)
over each boundary element,
$$
- S^0[\bargam_{ij},h_{ij}] = -S^0_\barT[\bargam_{ij}]
- S^0_{\Sigma''}[h_{ij}] + S^0_{\Sigma'}[h_{ij}]\, ,
$$
where $S^0_{\Sigma''}$ and $S^0_{\Sigma'}$ are the same functional.
The $\Sigma'$ and $\Sigma''$ contributions to $-S^0$ give rise to 
canonical transformation of the gravitational momentum $P^{ij}$ defined
in Eq.~(\ref{momenta}a). The $\barT$ contribution modifies 
$\bar{\Pi}^{ij}$ in Eq.~(\ref{momenta}b). Modification of 
$\bar{\Pi}^{ij}$ corresponds to the zero--point freedom in
$\bar{\varepsilon}$, $\barj_a$, and
$\bar{s}{}^{ab}$.

\vskip 3mm \noindent
14. See Ref.~\cite{Lau1} for the origins of
this transformation. It may also be found by carrying out the null
limit construction
described in Ref.~\cite{BLY1} for the particular case of
Schwarzschild.

\vskip 3mm \noindent
15. A similar analysis can be carried out for spacetimes which are
asymptotically anti-de Sitter\cite{HenTei}; and in particular,
by inclusion of a lapse factor into the energy surface integral, 
one can recover the Abbott-Deser definition energy \cite{AbbDes} 
in the $r\rightarrow\infty$ limit.\cite{BCM,HH1,Lau4}

\vskip 3mm \noindent
16. The detailed
connection between the quasilocal
energy and the ADM energy has not been derived previously in
the literature. See,
however, Refs.~\cite{BCM,HH1}.

\vskip 3mm \noindent
17. Be careful not to confuse
the metric $f_{ij}$ with the Euclidean metric of the auxiliary space 
$\referenceE^{3}$ used to obtain $k^0$, for the metric $f_{ij}$ lives  
on $\Sigma$.

\vskip 3mm \noindent
18. These {\em Ans\"{a}tze} can be
confirmed directly by computing $k_{ab}$ in linear  
approximation, a calculation which uses the fall-off
of $h_{ij}$ and the fact that $\partial_k r$ is proportional to
the one-form normal $n_{k}$ of $B$ in $\Sigma$.

\vskip 3mm \noindent
19. Ashtekar and Hansen\cite{AshHan}
define an expression at infinity via a conformal compactification. 
This integral is the physical spacetime version of the 
Ashtekar-Hansen expression.

\vskip 3mm \noindent
20. The four--velocity $U(t)$ is expressed in Minkowski
coordinates so that its components are unchanged when
$U(t)$ is parallel transported to
$t=0$.

\end{document}